\documentclass[aps,prd,10pt,a4paper,superscriptaddress,preprintnumbers,showpacs,notitlepage,eqsecnum,nofootinbib]{revtex4-1}

\usepackage[english]{babel}
\usepackage{graphicx}
\usepackage{amsmath}
\usepackage{xcolor}
\usepackage{eufrak}
\usepackage{caption}
\usepackage{subcaption}
\def\square{\mathchoice\sqr54\sqr54\sqr{2.1}3\sqr{1.5}3}
\def\sqr#1#2{{\vcenter{\vbox{\hrule height.#2pt\hbox{\vrule
width.#2pt height#1pt \kern#1pt\vrule width.#2pt}\hrule height.#2pt}}}}
\def\square{\mathchoice\sqr54\sqr54\sqr{2.1}3\sqr{1.5}3}

\makeatletter

\@addtoreset{equation}{section}
\makeatother

\begin{document}

\title{Two-body problem in Scalar-Tensor theories\\ as a deformation of General Relativity :\\an Effective-One-Body approach\\}

\author{F\'elix-Louis Juli\'e}
\author{Nathalie Deruelle}

\affiliation{APC, Universit{\'e} Paris Diderot,\\
CNRS, CEA, Observatoire de Paris, Sorbonne Paris Cit\'e\\
 10, rue Alice Domon et L{\'e}onie Duquet, F-75205 Paris CEDEX 13, France.}

\date{Wednesday March 15th, 2017}

\begin{abstract}
In this paper we address the two-body problem in massless Scalar-Tensor (ST) theories within an Effective-One-Body (EOB) framework. We focus on the first building block of the EOB approach, that is, mapping the conservative part of the  two-body dynamics onto the geodesic motion of a test particle in an effective external metric. To this end, we first deduce the second post-Keplerian (2PK) Hamiltonian of the two-body problem from the known 2PK Lagrangian. We then build, by means of a canonical transformation a ST-deformation of the general relativistic EOB Hamiltonian which allows to incorporate the Scalar-Tensor (2PK) corrections to the currently best available General Relativity EOB results. This EOB-ST Hamiltonian defines a resummation of the dynamics that  may provide information on the strong-field regime, in particular, the ISCO location and associated orbital frequency and can be compared to, other, e.g. tidal, corrections.
\end{abstract}

\maketitle

\section{Introduction\label{introduction}}
On  September 14th 2015, the two antennas of the ``Laser Interferometric Gravitational Observatory" (LIGO) detected a ``chirp" signal, called GW150914, which heralded a new era in gravitational wave astronomy. Indeed, up to then, only the back reaction of the emitted gravitational waves onto the inspiralling of binary stars had been observed, through pulsar timing, and found to be in full agreement with the general relativistic predictions, see, e.g., \cite{Hulse:1974eb} \cite{Stairs:2003eg} \cite{Possenti2016}. That time, an actual gravitational waveform was extracted from the data by the LIGO-Virgo collaboration, which was announced on February 11th 2016 to describe the first ever observed merger of two black holes \cite{Abbott:2016blz}. \\

Building libraries of accurate gravitational waveform templates is essential for the successful detection of the inspiral, merging and ``ring down" phases of binary systems of compact objects driven by gravity. In the framework of General Relativity (GR), post-Newtonian (PN) expansions of Einstein's equations are suitable to describe the weak field inspiral phase and the associated gravitational waveforms, and numerical relativity is required to take account of the full non-linear dynamics of the merging, whereas the settling down of the final black hole through its ringing modes can be tackled by semi-analytical methods, see, e.g., the reviews, \cite{Blanchet:2013haa}, \cite{Bishop:2016lgv}, \cite{Kokkotas1999}.\\

The ``Effective-One-Body" (EOB) approach has proven to be a very powerful way to analytically match and encompass the general relativistic post-Newtonian and numerical descriptions of the inspiralling and merging (as well as ring-down) phases of the dynamics of binary systems of comparable masses. It was initiated by A. Buonanno and T. Damour in 1998 \cite{Buonanno:1998gg} who reduced the general relativistic two-body problem at 2PN order\footnote{That is, up to and including $(v/c)^4$ corrections to the Newtonian dynamics.} to that of the geodesic motion of a test particle in an  effective external metric. They did so by mapping, by means of a canonical transformation, the two-body 2PN general relativistic Hamiltonian towards a much simpler, EOB Hamiltonian related to that of a test particle in geodesic motion in an external, static and spherically symmetric, metric.  Taking then this EOB Hamiltonian as exact (which amounts to an implicit resummation) and including the 2.5PN radiation reaction force, they described the inspiralling phase {\it up to}  merging, that is up to and through the last stable orbit. The gravitational waveforms thus predicted \cite{Buonanno:2000ef} turned out to be much simpler than previously argued \cite{Brady:1998du}, a simplicity which was confirmed later by numerical relativity \cite{Pretorius:2005gq}. This EOB approach was then extended to 3PN \cite{Damour:2000we} (see \cite{Damour:2016bks} for a review) and, recently, to 4PN \cite{Damour:2015isa}, and even beyond by introducing a couple of  parameters which are tuned by matching  to numerical relativity results. The corresponding gravitational waveforms were constructed using various resummation techniques, and were used to extract from the GW150914 data the characteristics of the coalescing black holes, that is their initial and final masses and spins, see e.g. \cite{Pan:2013tva}.\\

Matching binary system gravitational waveform templates to the present and forthcoming data from the LIGO-Virgo interferometers (and forthcoming detectors such as LISA) will allow to test gravity theories at high post-newtonian order (if the inspiralling phase can be monitored for a sufficient number of cycles) and in the strong field regime at merger. The template libraries being based at present on general relativistic waveforms, the tests and their feasibility are limited to phenomenological bounds on some parametrized PN coefficients, see e.g. \cite{TheLIGOScientific:2016src}, \cite{Yunes:2009ke}, \cite{Huwyler:2014gaa}.\\

A next step to test gravity theories in their strong field regimes using gravitational wave detectors is to match their data with templates predicted within the framework of ``modified gravities", that is, theories alternative to General Relativity. Among the large ``zoo" of modified gravities, Scalar-Tensor theories (ST) are probably those which are best motivated from a theoretical point of view and most studied. In their simplest versions, they consist in adding one massless scalar degree of freedom to gravity which, like the metric, couples universally to matter. They were introduced by Jordan, Fierz, Thiry, Brans and Dicke (see \cite{Goenner:2012cq} for a historical review) and were put in a modern perspective by Will and Zaglauer \cite{Will:1989sk}, Nordvedt \cite{Nordtvedt:1970uv}, and Damour and Esposito-Far\`ese \cite{Damour:1992we}. However the corresponding dynamics of binary systems is known at 2.5PN order only \cite{Mirshekari:2013vb}\footnote{or, in the terminology of \cite{Damour:1992we}, 2.5 post-Keplerian (2.5PK) order, to make  explicit the fact that it includes the scalar field dependence of the masses due to
self-gravity effects.\\} or numerically \cite{Shibata:2013pra}. What was hence done in \cite{Lang:2013fna}-\cite{Lang:2014osa} or \cite{Sennett:2016klh} is the computation of the gravitational waveforms in Scalar-Tensor theories at 2PN relative order.\\

What we propose to do here is a first step to go beyond what has been done up to now by extending to Scalar-Tensor theories the effective one body approach of Buonanno and Damour in their 1998 paper \cite{Buonanno:1998gg}.\\

 More precisely we will start from the Scalar-Tensor Lagrangian of two non-spinning bodies obtained by Mirshekari and Will in \cite{Mirshekari:2013vb}, which is the Scalar-Tensor extension of the Damour-Deruelle two-body GR Lagrangian \cite{DamourDeruelle} \cite{Damour2}. This ``Jordan-frame" Lagrangian, written in harmonic coordinates, depends on the positions, velocities and accelerations of the two bodies and we will first reduce it to an ordinary ``Einstein-frame" Lagrangian depending on  positions and velocities only by means of a contact transformation similar to what is done in GR \cite{schafer83}-\cite{Damour:1990jh}. It is then an exercise to obtain the centre-of-mass 2PK Hamiltonian.\\
 
 This 2PK Hamiltonian will then be mapped, after an appropriately chosen canonical transformation, to an effective one-body Hamiltonian, which reduces to the 1998 Buonanno-Damour EOB Hamiltonian \cite{Buonanno:1998gg} in the General Relativity limit. Of course the (conservative) dynamics derived from this EOB-ST Hamiltonian is the same at 2PK order as the dynamics derived from the Mirshekari-Will 2PN Lagrangian but, when taken as being exact, it defines an implicit resummation and hence different dynamics in the strong field regime which is reached near the last stable orbit. \\ 
 
 This  2PK EOB-ST Hamiltonian, which can be seen as a Scalar-Tensor ``deformation" of the general relativistic 2PN EOB one, will then be extended to incorporate the scalar-tensor 2PK corrections into the currently best available General Relativity EOB results,  and will thus be well-suited to test Scalar-Tensor theories when considered as parametrised corrections to General Relativity.\\
 
The paper is organised as follows : we first recall the general settings of Scalar-Tensor theories in section \ref{sectionSTReminder}.
In section \ref{sectionHamiltonian} we perform an order reduction of the conservative two-body Lagrangian at 2PK order. We then express the corresponding Hamiltonian in the centre-of-mass frame. In section \ref{sectionDroste} we build the effective problem, that is,
the geodesic motion in a 2PN effective external metric, and rewrite its dynamics as a ``ST-deformation" of General Relativity at 2PN order. We then incorporate the Scalar-Tensor 2PK corrections that we have obtained into the currently best available EOB-NR general relativistic Hamiltonian and, after adequate ``padeisation", some features of the corrections to the General Relativity ISCO predictions are described.

\section{Scalar-Tensor theory Reminder\label{sectionSTReminder}}
In this paper we adopt the conventions of Damour and Esposito-Far\`ese (see e.g. \cite{Damour:1992we} or \cite{Damour:1995kt}) and limit ourselves to the single, massless scalar field case. In the Einstein-frame, the action reads\footnote{For a comparison with the Jordan-frame parametrization of e.g. \cite{Mirshekari:2013vb}, see appendix \ref{appendixEvsJ}.\\}
\begin{equation}
S_{EF}=\frac{c^4}{16\pi G_{*}}\int d^4x\sqrt{-g}\bigg(R-2g^{\mu\nu}\partial_\mu\varphi\partial_\nu\varphi\bigg)+S_{m}\left[\Psi,\mathcal{A}^2 (\varphi) g_{\mu\nu} \right]\ ,
\label{actionEF}
\end{equation}
where $R$ is the Ricci scalar, $g\equiv \det g_{\mu\nu}$, and $\Psi$ generically stands for matter fields. In the following we shall work in units where $G_{*}\equiv 1$ and $c\equiv 1$ if not specified.
The free dynamics of the Einstein metric $g_{\mu\nu}$, which describes the tensorial degrees of freedom of gravity, is governed by the usual Einstein-Hilbert action. The dynamics of the scalar field $\varphi$, that is, the gravitational scalar degree of freedom of gravity, arises from its coupling to the matter fields $\Psi$. Indeed, matter minimally couples, not to the Einstein, but to the Jordan metric
\begin{equation}
\tilde g_{\mu\nu}\equiv \mathcal{A}^2(\varphi)g_{\mu\nu}\ .\label{defJordanMetric}
\end{equation}
This Jordan metric $\tilde g_{\mu\nu}$ is often referred to as the ``physical" one, since one retrieves Special Relativity in its locally inertial frames (i.e. frames where $\tilde g_{\mu\nu}=\eta_{\mu\nu}$, $\partial_\lambda\tilde g_{\mu\nu}=0$).
Hence, by construction, Scalar-Tensor theories explicitly encompass the Einstein Equivalence Principle \cite{DiCasola:2013iia}. A given Scalar-Tensor theory is completely determined once the function $\mathcal A(\varphi)$ has been specified. In particular, General Relativity is recovered for $\mathcal A(\varphi)=cst$.\\

From (\ref{actionEF}) one derives the Einstein-frame field equations :
\begin{subequations}
\begin{align}
&R_{\mu\nu}=2\partial_\mu\varphi\partial_\nu\varphi+8\pi\left(T_{\mu\nu}-\frac{1}{2}g_{\mu\nu}T\right)\ ,\\
&\square\varphi=-4\pi\alpha(\varphi)T\ ,
\end{align}
\label{EOFst}
\end{subequations}
where $R_{\mu\nu}$ is the Ricci tensor, $T_{\mu\nu}\equiv -\frac{2}{\sqrt{-g}}\frac{\delta S_m}{\delta g^{\mu\nu}}$ is the Eintein-frame energy-momentum tensor, $T\equiv T^\mu_{\phantom{\mu}\mu}$ and
\begin{equation}
\alpha(\varphi)\equiv \frac{d\ln \mathcal A(\varphi)}{d\varphi}
\end{equation}
measures the coupling between the scalar field and matter.\\

When dealing with compact, self-gravitating bodies (e.g. neutron stars or black holes), we adopt the phenomenological treatment suggested by Eardley \cite{Eardley} and justified by Damour \cite{lesHouches} and Damour and Esposito-Far\`ese \cite{Damour:1992we}, and ``skeletonize" these extended bodies as point particles :
\begin{equation}
S_m=-\sum_A\int d\lambda\sqrt{-\tilde{g}_{\mu\nu}\frac{dx^\mu}{d\lambda}\frac{dx^\nu}{d\lambda}}\tilde{m}_A(\varphi)\ ,
\end{equation}
where $\lambda$ is an affine parameter along the worldline of the particle. The Jordan-frame mass $\tilde{m}_A(\varphi)$ is not a constant but rather depends on the local value of the scalar field, on the specific theory and on body $A$ itself (through its equation of state in particular).\footnote{This scalar field dependence of the mass of point-like objects embodies the fact that the equilibrium configuration of an extended body depends on the value of the background scalar field at its location, imposed by the other (faraway) companions. We will not discuss here how this $\varphi$-dependence, or ``sensitivity" is (numerically) calculated, see e.g. \cite{Damour:1992we}, \cite{Damour:1993hw}, \cite{Zaglauer:1992bp}.\\}
 Since $\tilde{g}_{\mu\nu}=\mathcal A^2(\varphi)g_{\mu\nu}$, one also has
\begin{equation}
S_m=-\sum_A\int d\lambda\sqrt{-g_{\mu\nu}\frac{dx^\mu}{d\lambda}\frac{dx^\nu}{d\lambda}}m_A(\varphi)\ ,
\label{skeletonAction}
\end{equation}
where we have defined the Einstein-frame mass of the skeletonized compact bodies  as :
\begin{equation}
m_A(\varphi)\equiv \mathcal A(\varphi)\tilde{m}_A(\varphi)\ ,
\label{JFmassEF}
\end{equation}
that takes into account both the universal factor $\mathcal A(\varphi)$ and body-dependent self-gravity effects, $\tilde m_A(\varphi)$. Hence the two-body problem in ST theories is fully described by {\it two} functions, $m_A(\varphi)$  and $m_B(\varphi)$. The trajectory of freely falling bodies will generally no longer be universal anymore, thus violating the so-called Strong Equivalence Principle, unless their self-gravity is negligible, that is when $\tilde{m}_A$ and $\tilde{m}_B$ are constant (in which case they are geodesics of the Jordan metric).
In contrast, static, spherically symmetric black holes are known to carry no  massless scalar ``hair" and hence reduce to Schwarzschild black holes (see e.g. \cite{Hawking:1972qk} and \cite{Damour:1992we}). In that case $m_A(\varphi)=cst$, and black holes follow the geodesics of the Einstein metric. Consequently, binary black holes are usually expected to generate no deviation from General Relativity. However this is not guaranteed in the strong field, dynamical, regime of a binary coalescence, see conclusion.

Finally,  the following six, dimensionless,  body-dependent functions  built out of the two mass functions $m_A(\varphi)$ and $m_B(\varphi)$, will be useful at the 2PK order we are going to work at :
\begin{subequations}
\begin{align}
&\alpha_A(\varphi)\equiv\frac{d\ln m_A}{d\varphi}=\frac{d\ln \mathcal A}{d\varphi}+\frac{d\ln\tilde{m}_A}{d\varphi}\ ,\label{defSensitivitiesA}\\
&\beta_A(\varphi)\equiv\frac{d\alpha_A}{d\varphi}\ ,\label{defSensitivitiesB}\\
&\beta'_A(\varphi)\equiv\frac{d\beta_A}{d\varphi}\ .\label{defSensitivitiesC}
\end{align}
\label{defSensitivities}
\end{subequations}
In the negligible self-gravity limit, $\tilde{m}_A=cst$, these functions become universal :
\[
\alpha_A\rightarrow\alpha=\frac{d\ln \mathcal A}{d\varphi}\ ,\quad\beta_A\rightarrow\beta\equiv\frac{d\alpha}{d\varphi}\ ,\quad\beta'_A\rightarrow\beta'\equiv\frac{d\beta}{d\varphi}\ ,
\]
while in the General Relativity limit ($m_A(\varphi)=cst$), $
\alpha_A=\beta_A=\beta'_A=0$.

\section{The two-body 2PK conservative hamiltonian\label{sectionHamiltonian}}
The Scalar-Tensor two-body conservative Lagrangian has already been derived at second post-Keplerian order and will be our starting point. In particular, its structure (derived from a Fokker action) was given by Damour and Esposito-Far\`ese in \cite{Damour:1995kt} using a diagrammatic approach, while Mirshekari and Will provided its explicit expression in \cite{Mirshekari:2013vb}. Because of the harmonic coordinates in which it has been derived, this Lagrangian depends (linearly) on the accelerations of the bodies.

In this section we rewrite the Mirshekari-Will Lagrangian in the Einstein-frame conventions introduced above and in a class of coordinate systems where the Lagrangian is ordinary (i.e. only depends on positions and velocities). We then derive the associated Hamiltonian.
Finally, we transform it by means of a generic canonical transformation,  to prepare the mapping towards the effective problem that will be performed in section  \ref{sectionDroste}.

\subsection{Jordan-frame vs Einstein-frame}
From now on, any quantity that is related to the Jordan-frame will be denoted with a tilde superscript.
The Jordan-frame two-body Lagrangian has been derived at 2PK order in harmonic coordinates in \cite{Mirshekari:2013vb}, using a  set of Brans-Dicke-like parameters.
In order to rewrite it in terms of the Einstein-frame parametrization discussed above, one has to :

(i) translate the parameters of \cite{Mirshekari:2013vb} in terms of (\ref{defSensitivities}). The conversion is given in details in appendix \ref{appendixEvsJ} ;

(ii) note that the Jordan-frame Lagrangian of \cite{Mirshekari:2013vb} is written in a coordinate system $\{\tilde{x}^\mu\}$ such that the Jordan metric $\tilde{g}_{\mu\nu}\rightarrow\eta_{\mu\nu}$ is Minkowski at infinity, while in the Einstein-frame one uses instead coordinates $\{x^\mu\}$ such that $g_{\mu\nu}\rightarrow\eta_{\mu\nu}$. 
Since $\tilde{g}_{\mu\nu}=\mathcal A^2(\varphi)g_{\mu\nu}$,  that means the following global rescaling of coordinates has to be performed between both frames~:
\begin{equation}
\tilde{x}^{\mu}=\mathcal A_0x^\mu\ ,
\end{equation}
where, and from now on, a ``$0$" index indicates a quantity evaluated at $\varphi=\varphi_0$, where $\varphi_0$ is taken to be the asymptotic constant value of the scalar field far from the system, imposed by cosmology.
Therefore, in order to get the Einstein-frame Lagrangian, one has to  rescale the radial variable $R$ of \cite{Mirshekari:2013vb} to $\mathcal A_0 R$. For the same reasons, $t\rightarrow \mathcal A_0 t$ i.e. the Lagrangian has to be rescaled by an overall $\mathcal A_0$ factor.

All that taken into account, the Mirshekari-Will  two-body 2PK Lagrangian translates, in the Einstein-frame and in harmonic coordinates, as~:
\begin{equation}
L=-m_A^0-m_B^0+L_{\rm K}+L_{\rm 1PK}+L_{\rm 2PK}+\cdots
\label{Lagr2body}
\end{equation}
with
\begin{equation}
L_{\rm K}=\frac{1}{2}m_A^0V_A^2+\frac{1}{2}m_B^0V_B^2+\frac{G_{AB}m_A^0m_B^0}{R}\ ,
\end{equation}
 
\begin{align}
&L_{\rm 1PK}=\frac{1}{8}m_A^0V_A^4+\frac{1}{8}m_B^0V_B^4+\frac{G_{AB}m_A^0m_B^0}{R}\left(\frac{3}{2}(V_A^2+V_B^2)-\frac{7}{2}\vec{V}_A\cdot\vec{V}_B-\frac{1}{2}(\vec{N}\cdot\vec{V}_A)(\vec{N}\cdot\vec{V}_B)+\bar\gamma_{AB}(\vec{V}_A-\vec{V}_B)^2\right)\nonumber\\
&\hspace*{3cm}-\frac{G_{AB}^2m_A^0m_B^0}{2R^2}\bigg(m_A^0(1+2\bar{\beta}_B)+m_B^0(1+2\bar{\beta}_A)\bigg)\ ,
\end{align}
\begin{align}
&L_{\rm 2PK}=\frac{1}{16}m_A^0V_A^6\nonumber\\
&+\frac{G_{AB}m_A^0m_B^0}{R}\left[\frac{1}{8}(7+4\bar{\gamma}_{AB})\bigg(V_A^4-V_A^2(\vec{N}\cdot\vec{V}_B)^2\bigg)-(2+\bar{\gamma}_{AB})V_A^2(\vec{V}_A\cdot\vec{V}_B)+\frac{1}{8}(\vec{V}_A\cdot\vec{V}_B)^2\right.\nonumber\\
&\hspace*{2cm}\left.+\frac{1}{16}(15+8\bar{\gamma}_{AB})V_A^2V_B^2+\frac{3}{16}(\vec{N}\cdot\vec{V}_A)^2(\vec{N}\cdot\vec{V}_B)^2+\frac{1}{4}(3+2\bar{\gamma}_{AB})\vec{V}_A\cdot\vec{V}_B(\vec{N}\cdot\vec{V}_A)(\vec{N}\cdot\vec{V}_B)\right]\nonumber\\
&+\frac{G_{AB}^2m_B^0(m_A^0)^2}{R^2}\left[\frac{1}{8}\bigg(2+12\bar{\gamma}_{AB}+7\bar{\gamma}_{AB}^2+8\bar{\beta}_B-4\delta_A\bigg)V_A^2+\frac{1}{8}\bigg(14+20\bar{\gamma}_{AB}+7\bar{\gamma}_{AB}^2+4\bar{\beta}_B-4\delta_A\bigg)V_B^2\right.\nonumber\\
&\hspace*{1cm}-\frac{1}{4}\bigg(7+16\bar{\gamma}_{AB}+7\bar{\gamma}_{AB}^2+4\bar{\beta}_B-4\delta_A\bigg)\vec{V}_A\cdot\vec{V}_B-\frac{1}{4}\bigg(14+12\bar{\gamma}_{AB}+\bar{\gamma}_{AB}^2-8\bar{\beta}_B+4\delta_A\bigg)(\vec{V}_A\cdot\vec{N})(\vec{V}_B\cdot\vec{N})\nonumber\\
&\hspace*{1cm}\left.+\frac{1}{8}\bigg(28+20\bar{\gamma}_{AB}+\bar{\gamma}_{AB}^2-8\bar{\beta}_B+4\delta_A\bigg)(\vec{N}\cdot\vec{V}_A)^2+\frac{1}{8}\bigg(4+4\bar{\gamma}_{AB}+\bar{\gamma}_{AB}^2+4\delta_A\bigg)(\vec{N}\cdot\vec{V}_B)^2\right]\nonumber\\
&\left.+\frac{G_{AB}^3(m_A^0)^3m_B^0}{2R^3}\left[1+\frac{2}{3}\bar{\gamma}_{AB}+\frac{1}{6}\bar{\gamma}_{AB}^2+2\bar{\beta}_B+\frac{2}{3}\delta_A+\frac{1}{3}\epsilon_B\right]+\frac{G_{AB}^3(m_A^0)^2(m_B^0)^2}{8R^3}\bigg[19+8\bar{\gamma}_{AB}+8(\bar{\beta}_A+\bar{\beta}_B)+4\zeta\bigg]\right.\nonumber\\
&-\frac{1}{8}G_{AB}m_A^0m_B^0\bigg(2(7+4\bar{\gamma}_{AB})\vec{A}_A\cdot\vec{V}_B(\vec{N}\cdot\vec{V}_B)+\vec{N}\cdot\vec{A}_A(\vec{N}\cdot\vec{V}_B)^2-(7+4\bar{\gamma}_{AB})\vec{N}\cdot\vec{A}_AV_B^2\bigg)\nonumber\\
&\hspace*{1cm}+(A\leftrightarrow B)\ ,
\end{align}
where
\[\vec{N}=\frac{\vec{Z}_A-\vec{Z}_B}{R}\ ,\quad R=\mid \vec{Z}_A-\vec{Z}_B\mid\ ,\quad \vec{V}_A=\frac{d\vec{Z}_A}{dt}\ ,\quad \vec{A}_A=\frac{d\vec{V}_A}{dt}\ ,\]
$\vec{Z}_A$ being the position of particle A (in our system of units the radial coordinate $R$ has the dimension of a mass). As for the coefficients appearing in the two-body 2PK Lagrangian above, they are combinations of the following eleven constants, built out of the 8 functions defined in (\ref{defSensitivities}) when evaluated at infinity (all deduced, we recall, from the mass function $m_A(\varphi)$ and its $B$-counterpart which define the theory and bodies under study)~: \\
\begin{subequations}
\begin{align}
&m_A^0\ ,\quad  G_{AB}\equiv 1+\alpha_A^0\alpha_B^0\ ,\label{defParamPK1}\\
&\bar\gamma_{AB}\equiv -\frac{2\alpha_A^0\alpha_B^0}{1+\alpha_A^0\alpha_B^0}\ ,\quad
\bar\beta_A\equiv\frac{1}{2}\frac{\beta_A^0(\alpha_B^0)^2}{(1+\alpha_A^0\alpha_B^0)^2}\ ,
\label{defParamPK2}\\
&\delta_A\equiv\frac{(\alpha_A^0)^2}{(1+\alpha_A^0\alpha_B^0)^2}\ ,\quad\epsilon_A\equiv\frac{(\beta'_A\alpha_B^3)^0}{(1+\alpha_A^0\alpha_B^0)^3}\ ,\quad\zeta\equiv\frac{\beta_A^0\alpha_A^0\beta_B^0\alpha_B^0}{(1+\alpha_A^0\alpha_B^0)^3}\ ,
\label{defParamPK3}
\end{align}
\label{defParamPK}
\end{subequations}
(Our notations are a similar, yet simplified, version of the parameters introduced in \cite{Damour:1995kt} in the context of the $N$-body, multi-scalar problem and admit a diagrammatic interpretation, see \cite{Damour:1995kt}.)\footnote{In the post-Newtonian scheme, these parameters are expanded as series of the compactness $s\sim G_{*}m/c^2r$ of weakly self-gravitating bodies \cite{Damour:1995kt}.  In this paper the orbital velocity $\left(\frac{v}{c}\right)^2\sim\frac{G_{*}m}{c^2R}$ is the only perturbative parameter. Hence our ``post-Keplerian" (PK) scheme is valid even for strongly self-gravitating bodies.\\}
We note  that the effective (dimensionless, since we set $G_*=1$) gravitational constant $G_{AB}=1+\alpha_A^0\alpha_B^0$ \textit{does} depend on the bodies.\\

Although we shall stick to the Einstein-frame for the rest of this paper, the reader willing to rewrite any forthcoming result in terms of Jordan-frame variables should perform the replacements (we recall that tildes refer to the Jordan frame) :
\begin{align}
m_{A/B}^0\leftrightarrow\tilde m_{A/B}^0 \ , \quad G_{AB}\leftrightarrow\tilde{G}_{AB}\ , \quad R\leftrightarrow\tilde{R}\ , \quad \vec{N}\leftrightarrow\tilde{\vec{N}}\ ,\quad \vec{V}_{A/B}\leftrightarrow\tilde{\vec{V}}_{A/B}
\ ,\quad \vec{A}_{A/B}\leftrightarrow\tilde{\vec{A}}_{A/B}\ ,
\label{JFtoEF}
\end{align}
where
\begin{equation}
\tilde m_{A/B}^0=m_{A/B}^0/\mathcal A_0\ ,\quad \tilde{G}_{AB}=G_{AB}\mathcal A_0^2\ ,\quad\tilde{R}=\mathcal A_0R\ ,\quad\tilde{\vec{A}}_A=\vec{A}_A/\mathcal A_0 \quad\text{and}\quad \tilde{\vec{N}}=\vec{N}\ ,\quad\tilde{\vec{V}}_{A/B}=\vec{V}_{A/B}\ .
\label{JFtoEF2}
\end{equation}
As a final remark, the Lagrangian (\ref{Lagr2body}) generalizes the 2PN General Relativity one, obtained by Damour and Deruelle in harmonic coordinates in \cite{DamourDeruelle}, and reduces to it in the limit $m_A(\varphi)=cst$, i.e. 
\begin{align}
&\alpha_{A/B}=\beta_{A/B}=\beta'_{A/B}=0\\
\Rightarrow\hspace*{0,2cm} &G_{AB}=1\ ,\quad\bar\gamma_{AB}=\bar\beta_{A/B}=\delta_{A/B}=\epsilon_{A/B}=\zeta=0\ .
\label{limiteRG}
\end{align}

\subsection{The class of reduced Lagrangians\label{sectionElimTheAccel}}
The Lagrangian (\ref{Lagr2body}) is expressed in harmonic
 coordinates (i.e. such that 
 $\partial_\mu\left(\sqrt{-g}g^{\mu\nu}\right)=0$) and depends linearly on the accelerations $\vec A_A$ at the 2PK level. Let us add to it a (2PK) total time derivative, 
 \begin{equation}
 L\rightarrow L+\frac{df}{dt}\equiv L_f\ ,\label{Lf}
 \end{equation}
where $f$ is a generic function,
\begin{align}
&\frac{f}{m_A^0 m_B^0} \equiv G_{AB}\bigg[(f_1 V_A^2 + f_2 \vec V_A\cdot \vec V_B + f_3 V_B^2) (\vec N\cdot\vec V_A) - (f_4 V_A^2 + f_5 \vec V_A\cdot \vec V_B +
       f_6 V_B^2) (\vec N\cdot V_B)\cr
       & + f_7 (\vec N\cdot \vec V_A)^3 + f_8 (\vec N\cdot \vec V_A)^2 (\vec N\cdot \vec V_B) - 
   f_9 (\vec N\cdot\vec V_B)^2 (\vec N\cdot\vec V_A) - f_{10} (\vec N\cdot \vec V_B)^3\bigg]\cr
   & +G_{AB}^2\bigg[ f_{11} \left(\frac{m_A^0}{R}\right) (\vec N\cdot\vec V_A) + f_{12} \left(\frac{m_B^0}{R}\right) (\vec N\cdot\vec V_A) - 
   f_{13} \left(\frac{m_A^0}{R}\right) (\vec N\cdot \vec V_B) - f_{14} \left(\frac{m_B^0}{R}\right) (\vec N\cdot \vec V_B)\bigg]\ ,\label{functionF}
\end{align}
that depends on fourteen parameters (the $G_{AB}$ factor appears in the definition of the $f_i$ for dimensional convenience). This total derivative generates a boundary term and hence does not affect the equations of motion.\\

Now in order to deal with an ordinary Lagrangian (depending only on positions and velocities), a way to proceed is to reduce $L_f$ by ``boldly" replacing the accelerations by their leading order, that is Keplerian, on-shell expressions (as was done in \cite{OhtaEtAl} in General Relativity) :
\begin{subequations}
\begin{align}
L_f\rightarrow L_f\left(\vec{A}_A\rightarrow-\vec{N}\frac{G_{AB}m_B^0}{R^2}\ ,\ \vec{A}_B\rightarrow\vec{N}\frac{G_{AB}m_A^0}{R^2}\right)\equiv L_f^{red}
\end{align}
\label{ohta}
\end{subequations}
This indeed amounts to make an implicit 4-dimensional coordinate change, through a contact transformation, (see \cite{schafer83} \cite{Damour:1990jh}).
Hence the equations of motion derived from our reduced Lagrangian $L_f^{red}$ will be equivalent to those derived from \cite{Mirshekari:2013vb} but written in a different coordinate system, that depends on $f$. The full expression of the contact transformation is given  in appendix \ref{appendixTransfoContact}.\\

Hence we have on hand a whole class of coordinate systems (depending on the 14 parameters $f_i$) for which the class of Lagrangians $L_f^{red}$ is ordinary. The harmonic coordinates do not belong to that class.

\subsection{The centre-of-mass  two-body 2PK Hamiltonians\label{sectionHamilt2bod}}
We now derive the ordinary Hamiltonians, corresponding to the class of coordinate systems discussed above, by a further Legendre transformation,
\begin{align}
\vec{P}_A=\frac{\partial L_f^{red}}{\partial \vec{V}_A}\ , \quad\vec{P}_B=\frac{\partial L_f^{red}}{\partial \vec{V}_B}\ , \quad H=\vec{P}_A\cdot\vec{V}_A+\vec{P}_B\cdot\vec{V}_B-L_f^{red}\ .\label{HversusL}
\end{align}
In the centre-of-mass frame, $\vec{P}_A+\vec{P}_B\equiv\vec{0}$, and the conjugate variables are then easily checked to be $\vec{Z}\equiv \vec{Z}_A-\vec{Z}_B$ and $\vec{P}\equiv\vec{P}_A=-\vec{P}_B$. At 2PK order, when no spin effects come into play, the relative motion is planar. Hence, it is convenient to use polar coordinates $(R,\Phi)$, with conjugate momenta $P_R=(\vec{N}\cdot\vec{P})\hspace*{0,1cm},\hspace*{0,1cm}P_\Phi=R(\vec{N}\times\vec{P})_z$, setting $\theta=\pi/2$. From now on we denote $(Q,P)\equiv (R,\Phi,P_R,P_\Phi)$.\\

The general structure for an isotropic, translation-invariant, centre-of-mass frame, 2PK Hamiltonian $H(Q,P)$ is expected to be :
\begin{equation}
\hat{H}\equiv \frac{H}{\mu}=\frac{M}{\mu}+\left(\frac{\hat{P}^2}{2}-\frac{h^{\rm K}}{\hat{R}}\right)+\hat H^{\rm 1PK}+\hat H^{\rm 2PK}+\cdots
\label{ham2corpsStructure}
\end{equation}
with
\begin{subequations}
\begin{align}
&\hat H^{\rm 1PK} = \left(h^{\rm 1PK}_1 \hat P^4 + h^{\rm 1PK}_2\hat P^2\hat P_R^2 + 
   h^{\rm 1PK}_3 \hat P_R^4\right) +\frac{1}{\hat{R}}\left(h^{\rm 1PK}_4 \hat P^2 + h^{\rm 1PK}_5 \hat P_R^2\right) + \frac{h^{\rm 1PK}_6}{\hat R^2}\ ,\\ \nonumber \\
&\hat H^{\rm 2PK}=\left(h^{\rm 2PK}_1 \hat P^6+ h^{\rm 2PK}_2 \hat P^4 \hat P_R^2+h^{\rm 2PK}_3\hat P^2 \hat P_R^4 + h^{\rm 2PK}_4 \hat P_R^6\right) +\frac{1}{\hat R}\left(h^{\rm 2PK}_5 \hat P^4 +h^{\rm 2PK}_6 \hat P_R^2 \hat P^2 + h^{\rm 2PK}_7   \hat P_R^4\right)\nonumber\\
&\hspace*{1cm}+  \frac{1}{\hat R^2}\left(h^{\rm 2PK}_8\hat P^2 + h^{\rm 2PK}_9 \hat P_R^2 \right)+\frac{h^{\rm 2PK}_{10}}{\hat R^3}\ ,
\end{align}
\label{ham2corpsStructure2}
\end{subequations}
where we have introduced the dimensionless quantities
\begin{equation}
\hat P^2\equiv \hat P_R^2+{\hat P_\Phi^2\over \hat R^2}\quad\hbox{with}\quad \hat P_R\equiv\frac{P_R}{\mu}\ ,\ \hat P_\Phi\equiv{P_\Phi\over\mu M}\ ,\ \hat R\equiv {R\over M}\ ,
\label{impulsReduites}
\end{equation}
together with the reduced mass, total mass and symmetric mass ratio :
\begin{equation}
\mu\equiv{m_A^0m_B^0\over M}\ ,\ M\equiv m_A^0+m_B^0\ ,\ \nu\equiv{\mu\over M}\ .\label{reducedTotalMasses}
\end{equation}

The Scalar-Tensor Hamiltonians derived from the reduced Lagrangians (\ref{Lagr2body}), (\ref{Lf}-\ref{ohta}) fall into the class (\ref{ham2corpsStructure}-\ref{ham2corpsStructure2}), and their seventeen coefficients h$_i^{N{\rm PK}}$ are computed to be (written here when $f=0$ for simplicity)~: 

\begin{equation}
h^{\rm K}=G_{AB}\ ,\label{coeffsSTk}
\end{equation}
at Keplerian order,
\begin{align}
&h^{\rm 1PK}_1=-\frac{1}{8}(1-3\nu)\ ,\quad h^{\rm 1PK}_2=h^{\rm 1PK}_3=0\ ,\nonumber\\
h^{\rm 1PK}_4=-\frac{G_{AB}}{2}(3+\nu+2\bar\gamma_{AB})\ ,&\quad h^{\rm 1PK}_5=-\frac{G_{AB}}{2}\nu\ ,\quad h^{\rm 1PK}_6=\frac{G_{AB}^2}{2M}\bigg(m_A^0(1+2\bar\beta_B)+m_B^0(1+2\bar\beta_A)\bigg)\ ,\label{coeffsST1PK}
\end{align}
at 1PK order and
\begin{align}
&\hspace*{3cm}h^{\rm 2PK}_1=\frac{1}{16}\left(5 \nu^2-5 \nu+1\right),\quad h^{\rm 2PK}_2=h^{\rm 2PK}_3=h^{\rm 2PK}_4=0\ ,\nonumber\\
&h^{\rm 2PK}_5=\frac{G_{AB}}{8}\left[5  + 4 \bar{\gamma}_{AB} -( 22 + 16 \bar{\gamma}_{AB} ) \nu - 3  \nu^2\right]\ ,\quad
   h^{\rm 2PK}_6=-\frac{G_{AB}}{4} \nu (\nu - 1)\ , \quad
       h^{\rm 2PK}_7=-\frac{3G_{AB}}{8}\nu^2\ ,\nonumber\\ \nonumber\\
&h^{\rm 2PK}_8=\frac{G_{AB}^2}{8}\left[ 22 - 4 \frac{m_A^0\bar\beta_B+m_B^0\bar \beta_A }{M} + 4 \frac{m_A^0\delta_A +m_B^0\delta_B }{M} + 28 \bar\gamma_{AB} + 
   9 \bar\gamma_{AB}^2 + \nu\left(58 - 4\frac{m_A^0\bar \beta_A+m_B^0\bar\beta_B}{M} + 36 \bar\gamma_{AB}\right)\right]\nonumber\ ,\\
   &h^{\rm 2PK}_9=G_{AB}^2\left[-\frac{1}{2} - \frac{1}{2} \frac{m_A^0 \delta_A + m_B^0 \delta_B}{M} - \frac{\bar\gamma_{AB}}{2} - \frac{\bar\gamma_{AB}^2}{8} +\nu \left( 
   -4 + (\bar\beta_A + \bar\beta_B) - 3 \bar\gamma_{AB} + \frac{m_A^0\bar\beta_B+m_B^0\bar\beta_A}{M}\right)\right]\nonumber\ ,\\
  &h^{\rm 2PK}_{10}=G_{AB}^3
  \left[-\frac{1}{2}  - \frac{m_B^0\bar\beta_A+m_A^0\bar\beta_B}{M} - \frac{1}{6}\frac{m_A^0\epsilon_B+m_B^0\epsilon_A}{M}- \frac{1}{3}\frac{m_A^0\delta_A+m_B^0\delta_B}{M} - \frac{\bar\gamma_{AB}}{3}  - \frac{\bar\gamma_{AB}^2}{12}\right.\nonumber\ ,\\
  &\hspace*{2cm}+ 
   \nu \bigg(-\frac{15}{4} - \zeta + \frac{\bar\gamma_{AB}^2}{6} - \frac{4}{3} \bar\gamma_{AB} +  \frac{\delta_A + \delta_B}{3} + \frac{\epsilon_A+\epsilon_B}{6} -  (\bar\beta_A+\bar\beta_B)\bigg) \bigg]\ ,\label{coeffST2PK}
\end{align}
at 2PK order. 
\vfill\eject

The $f=0$ Scalar-Tensor two-body Hamiltonian given above is written in terms of the 17 coefficients  h$_i^{N{\rm PK}}$ which are in turn expressed in terms of the 11 constants (\ref{defParamPK}) (which are themselves functions of the 8 parameters $m_A^0$, $\alpha_A^0$, $\beta_A^0$ and $\beta_A^{'0}$ characterizing at 2PK order the functions $m_A(\varphi)$ and $m_B(\varphi)$).\footnote{\label{footnoteCoeffsNuls}The fact that $h_2^{1{\rm PK}}$, $h_3^{1{\rm PK}}$,  as well as $h_3^{2{\rm PK}}$, $h_3^{2{\rm PK}}$ and $h_4^{2{\rm PK}}$ vanish is due to the structure of the kinetic term, as will be seen in more detail below.\\}
In the other coordinate systems discussed in section \ref{sectionElimTheAccel}, the 14 coefficients of the function $f$ at 2PK order modify the ten 2PK coefficients, which can be found in appendix \ref{appendixHamiltonien2body}. Each $\{f_i\}$ setting implicitly corresponds to a distinct coordinate system.

\subsection{The canonically transformed class of real Hamiltonians\label{sectionTransfoCano}}
As discussed in the introduction, the EOB mapping requires imposing a functional relation between the ``real" two-body Hamiltonian $H(Q,P)$ (that is, the ST two-body 2PK class of Hamiltonians obtained in the previous subsection), and an effective Hamiltonian $H_e$, $H_e=f_{\rm EOB}(H)$, by means of a canonical transformation.

We thus perform a further general canonical transformation on the real two-body Hamiltonians $H(Q,P)$,
\begin{equation}
(Q,P)\rightarrow(q,p)\ ,
\end{equation}
where, for the moment, $(q,p)\equiv(r,\phi,p_r,p_\phi)$ is a distinct set of canonical variables with no particular interpretation.
The canonical transformation is generated by a time-independent function\footnote{The generating function cannot depend on time since it has to relate two conservative problems.\\} 
$F(q, Q)$ such that the Lagrangian is shifted by a total derivative $L_f^{red}(Q,\dot Q)=L'(q,\dot q)+dF/dt$ and the Hamiltonian is a scalar $H(Q,P)=H'(q,p)$, so that, see, e.g,  (\ref{HversusL})~:
\begin{align*}
S&\equiv\int\,(L' dt+dF)=\int(p_r\,dr+p_\phi\,d\phi-H dt+dF)\cr
&=\int(P_R\,dR+P_\Phi\,d\Phi-H dt)\quad\hbox{and thus}\quad dF=P_R\,dR+P_\Phi\,d\Phi-(p_r\,dr+p_\phi\,d\phi) \,.
\end{align*}
For practical reasons, we shall rather consider the generating function $G(Q,p)$ such that :
\begin{align}
G&\equiv F+(p_r\,r+p_\phi\,\phi)-(p_r\,R+p_\phi\,\Phi)\ ,\\
\Rightarrow\quad dG(Q,p)&=dR\left(P_R-p_r\right)+d\Phi\left(P_\Phi-p_\phi\right)+dp_r\left(r-R\right)+dp_\phi\left(\phi-\Phi\right)\ ,\nonumber
\end{align}
that leads to the canonical transformation
\begin{equation}
r(Q,p)=R+{\partial G\over\partial p_r}\ ,\quad \phi(Q,p)=\Phi+{\partial G\over\partial p_\phi}\ ,\quad P_R(Q,p)=p_r+{\partial G\over\partial R}\ ,\quad P_\Phi(Q,p)=p_\phi+{\partial G\over\partial\Phi}\ .
\label{transfoCano}
\end{equation}

We now consider a generic ansatz for $G$, that generates 1PK and higher order coordinate changes\footnote{We know from Newton's theory  that once written in the center-of-mass frame, the Keplerian two-body Hamiltonian does not necessitate any further canonical transformation  and is  the effective-one-body Hamiltonian. From (\ref{transfoCano}), one checks that the Keplerian order coordinate change is indeed the identity.\\}, which depends on nine parameters :
\begin{equation}
{G(Q,p)\over \mu M}=\hat R\, \hat p_r\left(\alpha_1{\cal P}^2+\beta_1\hat p_r^2+{\gamma_1\over\hat R}+\alpha_2{\cal P}^4+\beta_2{\cal P}^2\hat p_r^2+\gamma_2\hat p_r^4+\delta_2{{\cal P}^2\over \hat R}+\epsilon_2{\hat p_r^2\over \hat R}+{\eta_2\over \hat R^2}+\cdots\right)\ ,\label{generatrice}
\end{equation}
where we have introduced the dimensionless quantities
\begin{equation}
\quad {\cal P}^2\equiv\hat p_r^2+{\hat p_\phi^2\over\hat R^2}\ ,\quad \hat R\equiv {R\over M}\ ,\quad\hat p_r\equiv\frac{p_r}{\mu}\ ,\quad \hat p_\phi\equiv\frac{p_\phi}{\mu M}\ .
\end{equation}
We chose this generating function not to depend on $\Phi$ so that $P_\Phi=p_\phi$. Also, for circular orbits for which $p_r=0\Leftrightarrow P_R=0$, we note that $\phi=\Phi$ and hence only the radial coordinates differ $r\neq R$.\\

Rather than inverting iteratively both first relations of (\ref{transfoCano}),  the real and effective Hamiltonians will be expressed in the following in the intermediate coordinate system $(Q,p)$ for computational convenience.
The two-body Hamiltonian (\ref{ham2corpsStructure}-\ref{ham2corpsStructure2}), together with its coefficients (\ref{coeffsSTk}-\ref{coeffST2PK}), is transformed to the intermediate coordinate system $H'(Q,p)=H(Q,P)$ using the last two relations in (\ref{transfoCano}) and is computed to be
\begin{equation}
\hat H=\frac{M}{\mu}+\left(\frac{{\cal P}^2}{2}-\frac{h^{\rm K}}{\hat R}\right)+\hat H^{\rm 1PK}+\hat H^{\rm 2PK}+\cdots\ ,
\end{equation}
where $h^{\rm K}=G_{AB}$, and where the explicit expressions for $\hat H^{\rm 1PK}$ and $\hat H^{\rm 2PK}$ for a generic function $f$ are given in appendix \ref{appendixPostTranfoCano}. It depends on the 8 fundamental parameters characterizing the theory and the two-bodies at 2PK order, that is, $m_A^0$, $\alpha_A^0$, $\beta_A^0$, ${\beta'_A}^0$ and their $B$-counterparts, on the 14 parameters $f_i$ characterizing the coordinate system used, and the 9 parameters of the canonical transformation.

\subsection{The functional relation between the real and EOB Hamiltonians\label{sectionfunctionalRelation}}

We have obtained a class of ordinary 2PK Hamiltonians that implicitly correspond to different coordinate systems, $H(Q,P)$. By means of a canonical transformation we have transformed them into an even larger class $H(Q,p)$. Our aim in the next section will be to find the canonical transformations which relate them to the Hamiltonian $H_e$ of an effective-one-body problem by means of a functional relation, $H_e=f_{\rm EOB}(H)$.

At 2PK order, this functional relation can a priori be expanded as, substracting the rest-mass constants :
\begin{equation}
\frac{H_e(Q,p)}{\mu}-1=\left(\frac{H(Q,p)-M}{\mu}\right)
\left[1+\frac{\bar\nu_1}{2}\left(\frac{H(Q,p)-M}{\mu}\right)+\bar\nu_2\left(\frac{H(Q,p)-M}{\mu}\right)^2+\cdots\right]\ ,
\end{equation}
with the Hamiltonians identifying at Keplerian order.
Now, as  justified in detail in e.g. \cite{Buonanno:1998gg}, \cite{Damour:2000we} and  \cite{Damour:2015isa} up to at least  4PN in General Relativity, and as proven to be true at all orders in GR {\it as well as} in ST theories in  \cite{Damour:2016gwp} within a post-Minkowskian scheme, the relation must be quadratic \textit{at all orders}, with  $\bar\nu_1=\nu=\mu/M$, $\bar\nu_2=0\cdots$, that is
\begin{equation}
\frac{H_e(Q,p)}{\mu}-1=\left(\frac{H(Q,p)-M}{\mu}\right)
\left[1+\frac{\nu}{2}\left(\frac{H(Q,p)-M}{\mu}\right)\right]\ .
\label{EOBquadrRel}
\end{equation}
As we shall see, $H_e$ will be \textit{uniquely} determined. Inverting (\ref{EOBquadrRel}) hence defines the unique, ``resummed" EOB Hamiltonian :
\begin{equation}
H_{\rm EOB}=M\sqrt{1+2\nu\left(\frac{H_e}{\mu}-1\right)}\ .
\label{relRacine}
\end{equation}
The dynamics deduced from $H_{\rm EOB}$ and the ``real" Hamiltonians $H$ are, by construction, equivalent up to 2PK order.\\

The topic  of the next section \ref{sectionDroste} is to propose a Scalar-Tensor effective one body Hamiltonian $H_{\rm EOB}$ which reduces, in the limit where the scalar interaction is switched off, to the EOB Hamiltonian of General Relativity obtained in \cite{Buonanno:1998gg}.

\section{\textbf{ST-deformation of the general relativistic EOB Hamiltonian}\label{sectionDroste}}

In this section,  which is the core of the paper, we first recall the structure of the Hamiltonian $H_e$ for geodesic motion in an (effective) static, spherically symmetric metric (in Schwarzshild-Droste coordinates). We then proceed to the EOB mapping and show that the resulting effective metric is unique and can be considered as a Scalar-Tensor-deformed version of the 2PN results of \cite{Buonanno:1998gg}. 

\subsection{The 2PN geodesic dynamics in an effective external one body metric\label{sectionEffectGeodDroste}}
Let us consider a static, spherically symmetric metric, written in Schwarzshild-Droste coordinates (for $\theta=\pi/2$)~:
\begin{equation}
ds^2_e=-A(r)dt^2+B(r)dr^2+r^2d\phi^2\ .\label{effectiveGeodesic}
\end{equation}
The geodesic dynamics of a test particle coupled to this external metric, with mass $\mu$ (which is identified to the real two-body reduced mass defined in (\ref{reducedTotalMasses})),  is described by the Lagrangian
\begin{equation}
L_e=-\mu \sqrt{-g^e_{\mu\nu}\frac{dx^\mu}{dt}\frac{dx^\nu}{dt}}=-\mu\sqrt{A-B\dot r^2-r^2\dot\phi^2}
\end{equation}
where $\dot r\equiv dr/dt$, $\dot\phi\equiv d\phi/dt$. The (dimensionless) effective Hamiltonian is : 
\begin{align}
&\hat H_e\equiv \frac{H_e}{\mu}=\sqrt{A\left(1+{\hat p_r^2\over B}+{\hat p_\phi^2\over \hat r^2}\right)}\quad\hbox{with}\quad p_r\equiv{\partial L_e\over\partial\dot r}\quad,\quad p_\phi\equiv{\partial L_e\over\partial\dot\phi}\ ,
\label{hamilGeod}\\
&\text{and where}\quad\hat r\equiv{r\over M}\ ,\quad\hat p_r\equiv{p_r\over \mu}\ ,\quad\hat p_\phi\equiv{p_\phi\over \mu M}\ ,\quad\hat p^2\equiv\hat p_r^2+{\hat p_\phi^2\over \hat r^2}\ ,
\label{varChap1}
\end{align}
$M$ being an effective mass,  identified with the real two-body total mass.

Now, the  $A$ and $B$ metric functions are generically expanded (at the required 2PK order) as :
 \begin{equation}
 A(r)=1+\frac{a_1}{\hat{r}}+\frac{a_2}{\hat{r}^2}+\frac{a_3}{\hat{r}^3}+\cdots\ ,\quad B(r)=1+\frac{b_1}{\hat{r}}+\frac{b_2}{\hat{r}^2}+\cdots\ ,\label{abMetricCoeff}
 \end{equation}
where $a_1$, $a_2$, $a_3$, $b_1$ and $b_2$ are the 5 (dimensionless) effective parameters to be determined. The 2PN effective Hamiltonian then becomes
 \begin{equation}
 \hat{H}_e=1+\hat H^{\rm N}_e+\hat H^{\rm 1PN}_e+\hat H^{\rm 2PN}_e+\cdots\ ,\label{HamEffectDroste2PN}
 \end{equation}
 where
 \begin{align*}
  &\hat H^{\rm N}_e=\frac{\hat p^2}{2}+\frac{a_1}{2\hat{r}}\ ,\quad\hat H^{\rm 1PN}_e=-\frac{\hat{p}^4}{8}-\hat{p}_r^2\frac{b_1}{2\hat{r}}+\frac{1}{4}\frac{a_1}{\hat{r}}\hat{p}^2+\frac{a_2-a_1^2/4}{2\hat{r}^2}\ ,\\ \\
  &\hat H^{\rm 2PN}_e=\frac{\hat p^6}{16}-\frac{\hat p^4 a_1-4\hat p_r b_1}{16\hat r}+\frac{(4a_2-a_1^2)\hat p^2+4(2b_1^2-2b_2-a_1b_1)\hat p_r^2}{16\hat r^2}+\frac{a_1^3-4a_1a_2+8a_3}{16\hat r^3}\ .
 \end{align*}
 
In the previous section we performed a generic canonical transformation $(Q,P)\to(q,p)$ and wrote the real Hamiltonians $H(Q,P)$ in terms of the intermediate coordinates $(Q,p)$. In order to be in a position to relate them to the effective Hamiltonian $H_e(q,p)$ considered here, we have to express the latter in the same variables $(Q,p)$ by means of the generic canonical transformation (\ref{transfoCano}-\ref{generatrice}).
We thus turn it into the class of Hamiltonians (recalling the notation ${\cal P}^2\equiv\hat p_r^2+{\hat p_\phi^2/\hat R^2}$)~:
\begin{equation}
\hat H_e=\frac{M}{\mu}+\left(\frac{{\cal P}^2}{2}-\frac{h^{\rm K}}{\hat R}\right)+\hat H_e^{\rm 1PK}+\hat H_e^{\rm 2PK}+\cdots\ ,
\end{equation}
where the explicit expressions for $\hat H_e^{\rm 1PK}$ and $\hat H_e^{\rm 2PK}$ are given in appendix E. These Hamiltonians depend on the 5 parameters $a_i$ and $b_i$ entering the effective metric coefficients at 2PN order, see (\ref{abMetricCoeff}), and on the 9 parameters entering the canonical transformation (\ref{transfoCano}-\ref{generatrice}).
 
\subsection{The Scalar-Tensor effective one-body metric at 2PK order\label{EOBdroste}}

As we saw in section \ref{sectionfunctionalRelation} the effective Hamiltonian $H_e$ and  the two-body 2PK Hamiltonians $H$ obtained in section III must be related through the quadratic relation (\ref{EOBquadrRel}), that is :
$$
\frac{H_e(Q,p)}{\mu}-1=\left(\frac{H(Q,p)-M}{\mu}\right)
\left[1+\frac{\nu}{2}\left(\frac{H(Q,p)-M}{\mu}\right)\right]\ 
$$

Consider now the generic (theory-agnostic) two-body Hamiltonian written in terms of the 17 coefficients $h_i^{N\rm PK}$,  see (\ref{ham2corpsStructure}), (\ref{ham2corpsStructure2}). It turns out that an effective $H_e$ can be constructed at 1PK level provided that~:
\begin{equation}
2h^{\rm 1PK}_2+3h^{\rm 1PK}_3=0\,.
\end{equation}
Any theory (such as Scalar-Tensor) whose purely kinetical terms take the form $m^0_A\sqrt{1-V_A^2}+m^0_B\sqrt{1-V_B^2}$ at the Lagrangian level, is such that $h^{\rm 1PK}_2=0$ and $h^{\rm 1PK}_3=0$ (as anticipated in footnote \ref{footnoteCoeffsNuls}). Thus this condition is not restrictive. At 2PK level, the identification requires two further conditions ; the first one
\begin{equation}
h^{\rm 2PK}_4 = -{2\over45} \left(12 h^{\rm 2PK}_2 + 18h^{\rm 2PK}_3+(h^{\rm 1PK}_2)^2 \right)\end{equation}
is no more restrictive than (IV.8), for the same reasons. The second condition however
\begin{align}
&h^{\rm 2PK}_1 + {7\over3}h^{\rm 2PK}_2 + h^{\rm 2PK}_3 + h^{\rm 2PK}_5 + h^{\rm 2PK}_6 + 
  h^{\rm 2PK}_7 =\cr
  &-{h^{\rm K}\over128} (5 + 2 \nu + 5 \nu^2) + 
  {1\over8} (1 + \nu) \bigg( (3 h^{\rm 1PK}_1 + h^{\rm 1PK}_2)h^{\rm K} +  h^{\rm 1PK}_4 + h^{\rm 1PK}_5\bigg)+ 
  {5\over2} h^{\rm 1PK}_1 \bigg(7 h^{\rm 1PK}_1 h^{\rm K} + 2 ( h^{\rm 1PK}_4 + h^{\rm 1PK}_5)\bigg)\ \cr
  & + 
  {1\over6} h^{\rm 1PK}_2 \bigg(13 h^{\rm 1PK}_2h^{\rm K} + 10 ( h^{\rm 1PK}_4 + h^{\rm 1PK}_5)\bigg) + {35\over3} h^{\rm 1PK}_1 h^{\rm 1PK}_2 h^{\rm K}\ ,
\end{align}
\textit{is} restrictive and the mapping of the two-body problem towards an effective geodesic is only possible for a subclass of theories.

In the Scalar-Tensor case, one checks that the coefficients (\ref{coeffsSTk}-\ref{coeffST2PK}) (see also appendix \ref{appendixHamiltonien2body}) {\it do} satisfy the condition (IV.10) {\it whatever} the values of the 14 $f_i$ parameters, that is, independently of the coordinate system in which the two-body Hamiltonian has been written, as it should.\footnote{The relation (IV.10) is thus also verified by General Relativity. In contrast, it is not satisfied by Electrodynamics at second post-Coulombian order (see \cite{Buonanno:2000qq}).

It is an exercise to extend this computation to 3PK. The identification then requires a further condition at 2PK, which first is not natural and, second, is not satisfied by the ADM Hamiltonian of General Relativity \cite{Buonanno:1998gg}. In consequence, this condition will not be satisfied in Scalar-Tensor theories either (since they include GR as a limit) and, as in GR  \cite{Damour:2000we},  it will no longer be possible to  map the  two body problem towards a geodesic.\label{footnote3PN}\\}\\

Inserting now in the functional relation (\ref{EOBquadrRel}) recalled above the explicit expressions for the ST coefficients $h_i^{N\rm PK}$ of the real two-body Hamiltonians $H$ obtained in the previous section, the identification  term by term is then seen to yield a \textit{unique} solution for $H_e$ and hence for the effective one-body metric, which can be written as~:
\begin{align}
 &A(r)=1-2\left(\frac{G_{AB}M}{r}\right)+2\bigg[\langle\bar\beta\rangle-\bar\gamma_{AB}\bigg]\left(\frac{G_{AB}M}{r}\right)^2+\bigg[2\nu+\delta a_3^{\rm ST} \bigg]\left(\frac{G_{AB}M}{r}\right)^3+\cdots\ ,\label{AeffectDroste}\\
 &B(r)=1+2 \bigg[1 + \bar\gamma_{AB}\bigg]\left(\frac{G_{AB}M}{r}\right)+\bigg[2(2-3\nu)+\delta b_2^{\rm ST}\bigg]\left(\frac{G_{AB}M}{r}\right)^2+\cdots\ ,\label{BeffectDroste}
 \end{align}
 where
 \begin{align}
 &\delta a_3^{\rm ST}\equiv\frac{1}{12} \bigg[-20 \bar\gamma_{AB} - 35 \bar\gamma_{AB}^2 - 24  \langle\bar\beta\rangle(1 - 2 \bar\gamma_{AB}) + 
   4 \big(\langle\delta\rangle- \langle\epsilon\rangle\big)\label{deltaA3}\\
   & \hspace*{0,5cm}+ 
   \nu\bigg(- 36 (\bar\beta_A + \bar\beta_B)+ 4 \bar\gamma_{AB}(10 +  \bar\gamma_{AB}) +4 (\epsilon_A + \epsilon_B) + 8 (\delta_A + \delta_B) - 
      24 \zeta \bigg)\bigg]\ ,\nonumber \\
 &\delta b_2^{\rm ST}\equiv\left[4\langle\bar\beta\rangle- \langle\delta\rangle+ \bar\gamma_{AB}(9 + \frac{19}{4} \bar\gamma_{AB})+\nu\bigg(2 \langle\bar\beta\rangle-4 \bar\gamma_{AB}\bigg) \right]\ ,\label{deltaB2}
 \end{align}
where we introduced the ``mean" quantities
\begin{align}
\langle\bar\beta\rangle\equiv \frac{m_A^0\bar\beta_B+m_B^0\bar\beta_A}{M}\ ,\quad\langle\delta\rangle\equiv\frac{m_A^0 \delta_A + m_B^0 \delta_B}{M}\ ,\quad\langle\epsilon\rangle\equiv\frac{m_A^0 \epsilon_B + m_B^0 \epsilon_A}{M}\ .\label{RappelmeanOverBodies}
\end{align}
That is the main result of this paper, which shows that one can interpret 2PK Scalar-Tensor theories as a deformation of the 2PN General Relativity results \cite{Buonanno:1998gg} (which is retrieved in the limit (\ref{limiteRG})).

\vfill\eject

A few comments are in order~: \\ \\
(i) one sees that the bare (dimensionless) gravitational constant is replaced by the effective one $G_*\rightarrow G_{AB}$ at every order~;\footnote{As was done up to 2PK, the parameters (\ref{defParamPK}) can always be defined so as to factorize out the appropriate $(G_{AB})^n$ factor corrsponding to any $R^n$ term at the Lagrangian level (\ref{Lagr2body}). One may then anticipate this property to hold at higher PK orders, so that the coordinate $R$ always comes in the form $R/G_{AB}$.\\}\\ \\
(ii) one recognizes, at 1PK level,  a parametrized post-Newtonian Eddington metric written in Droste coordinates, with~:
\begin{equation*}
\beta^{\rm Edd}=1+\langle\bar\beta\rangle\ ,\quad \gamma^{\rm Edd}=1+\bar\gamma_{AB}
\end{equation*}
being the Eddington parameters, such that $\beta^{\rm Edd}=\gamma^{\rm Edd}=1$ in General Relativity. Interestingly, these effective Eddington parameters encompass the self-gravity of both real bodies through the simple mean quantities (\ref{RappelmeanOverBodies}), extending the results of \cite{Damour:2016gwp}.\footnote{We thank Thibault Damour for having pointed out to us this important feature.\\}
The reader should note that the two-body parameters (\ref{defParamPK}) were initially defined in \cite{Damour:1992we} consistently with the parametrized-post-Newtonian (PPN) approach, where the N-body problem is to be interpreted as point particles following geodesics of a PPN \textit{metric}. Hence it is not surprising that properties (i) and (ii) emerge in the context of a \textit{metric} effective problem.
 \\

It must also be noted that the effective metric does not depend on the function $f$ introduced in section \ref{sectionElimTheAccel}, i.e. on the coordinate system ($R$, $\Phi$) in which the two-body Hamiltonian is written. Indeed, as it should, $f$ is absorbed by the canonical transformation (\ref{transfoCano}-\ref{generatrice}), whose parameters are found to be :
\begin{align}
&\alpha_1=-\frac{\nu}{2}\ ,\quad\beta_1=0\ ,\quad\gamma_1=G_{AB} \left(1+\bar\gamma_{AB}+\frac{\nu}{2}\right)\ ,\quad\alpha_2=\frac{1}{8} (1 - \nu) \nu\ ,\quad\beta_2=0\ ,\quad\gamma_2=\frac{\nu^2}{2}\ ,\nonumber\\
&\delta_2=G_{AB}\left[f_6 \frac{m_A^0}{M} + f_1 \frac{m_B^0}{M}-\nu \left( f_1 + f_6+(-f_3 + f_5 + f_6)\frac{m_A^0}{M} + (f_1 + f_2-f_4 ) \frac{m_B^0}{M} - \frac{3}{2}(1+\bar\gamma_{AB})+\frac{\nu}{8}\right)\right]\ ,\nonumber\\
&\epsilon_2=G_{AB}\left[-\frac{\nu^2}{8} + f_{10} \frac{m_A^0}{M} + f_7 \frac{m_B^0}{M} - 
 \nu \left(  f_7+f_{10} + (f_9 + f_{10})\frac{m_A^0}{M} + (f_7 + f_8)\frac{m_B^0}{M}\right)\right]\ ,\nonumber\\
&\eta_2=\frac{1}{8} G_{AB}^2 \left[8 \langle\bar\beta\rangle - 4 \langle\delta\rangle + 4 \bar\gamma_{AB} + 3 \bar\gamma_{AB}^2 + 
    \nu \bigg(-38 + 4 (\bar\beta_A + \bar\beta_B) - 24\bar\gamma_{AB} + 2  \nu\bigg)\right] \nonumber\\
    &\hspace*{2cm} +G_{AB}^2\bigg( 
    f_{13} \frac{m_A^0}{M}+f_{12} \frac{m_B^0}{M} + \nu ( f_{11} - f_{12} - f_{13} + f_{14} )\bigg)\ ,\label{coeffsCanoDroste}
\end{align}
and reduce to the General Relativity (ADM) values of \cite{Buonanno:1998gg} in the limit (\ref{limiteRG}), and (\ref{limiteADM}).\\

In this section, the Scalar-Tensor two-body problem has thus been mapped towards the geodesic of an effective external metric (\ref{effectiveGeodesic}), written in Droste coordinates, which is well-suited  when Scalar-Tensor effects are to be considered as perturbative with respect to General Relativity. We turn in the next subsection to the study of some aspects of the dynamics this EOB problem defines.

\subsection{The 2PK effective problem as a ST-deformation of General Relativity at  2PN order\label{sectionParametrizedEOB}}
Solar system and binary pulsar observations have put stringent constraints on Scalar-Tensor theories. In particular, the decay rate of the orbital period of binary pulsars (excluding dipolar radiation)  has led to the constraint (see \cite{Antoniadis:2013pzd}, \cite{Freire:2012mg}) :
\begin{equation}
(\alpha_A^0)^2<4\times 10^{-6}\ ,\label{contrainteBinaryPulsars}
\end{equation}
for any body $A$, regardless of its self-gravity or equation of state.\footnote{In particular,  this bound constrains certain classes of Scalar-Tensor theories which predict that strongly self-gravitating bodies such as neutron stars can develop significant scalar ``charges", i.e. a significant $\alpha_A^0$ parameter, even when $\alpha^0=\left.\frac{d\ln\mathcal A}{d\varphi}\right|_{\varphi_0}$ is vanishingly small, see \cite{Damour:1993hw}.} Now, the two-body Lagrangian parameters (\ref{defParamPK}) are all driven by a factor $(\alpha_{A/B}^0)^i$, where $i\geq 2$ (as can be understood from the diagrammatic approach of \cite{Damour:1995kt}) and can be conjectured to be all of the same order.
This overall factor is also seen to appear  at the level of the Scalar-Tensor corrections to $A(r)$ and $B(r)$ at any PK order, see (\ref{AeffectDroste}-\ref{deltaB2}).

\vfill\eject

Hence, the dynamics defined by the effective metric (\ref{effectiveGeodesic}), that is
\begin{equation}
ds_e^2=-A(r)dt^2+B(r)dr^2+r^2d\phi^2\ ,\label{geodEffDroste}
\end{equation}
with $A$ and $B$ given in (\ref{AeffectDroste}-\ref{deltaB2}), is particularly well-suited to regard Scalar-Tensor effects as perturbations to General Relativity. Remarkably, and as we shall recall below, when studying the conservative dynamics of circular orbits in Droste coordinates, only the $g^e_{00}$ component of the metric intervenes and can be written as
 \begin{equation}
 A(r)=A^{\rm GR}_{2\rm PN}\left(\frac{G_{AB}M}{r} ; \nu\right)+\delta A^{\rm ST}\left(\frac{G_{AB}M}{r} ; \nu\right)\ ,\label{AplusCorrections}
 \end{equation}
where $A^{\rm GR}_{2\rm PN}\left(\frac{G_{AB}M}{r} ; \nu\right)$ is the 2PN GR limit obtained by Buonanno-Damour (with $M\to G_{AB}M$) and where, as can be seen from (IV.11-13)~:
\begin{subequations}
\begin{align}
&\delta A^{\rm ST}=\delta A^{\rm ST}_{\rm 1PK}+\delta A^{\rm ST}_{\rm 2PK}\ ,\\
\delta A^{\rm ST}_{\rm 1PK}(r)=2\left(\frac{G_{AB}M}{r}\right)^2&\bigg[\langle\bar\beta\rangle-\bar\gamma_{AB}\bigg]\ ,\quad\delta A^{\rm ST}_{\rm 2PK}(r)=\left(\frac{G_{AB}M}{r}\right)^3\delta a_3^{\rm ST}(\nu)\ .
\end{align}
\end{subequations}
Introducing finally the notations
\begin{align}
&\frac{G_{AB}M}{r}\equiv u\  ,\quad\langle\bar\beta\rangle-\bar\gamma_{AB}\equiv\epsilon_{\rm 1PK} ,\quad \delta a_3^{\rm ST}(\nu)\equiv\epsilon^0_{\rm 2PK}+\nu\,\epsilon^\nu_{\rm 2PK}\ ,\\
\text{where}\quad &\epsilon^0_{\rm 2PK}\equiv\frac{1}{12} \bigg[-20 \bar\gamma_{AB} - 35 \bar\gamma_{AB}^2 - 24  \langle\bar\beta\rangle(1 - 2 \bar\gamma_{AB}) + 
   4 \big(\langle\delta\rangle- \langle\epsilon\rangle\big)\bigg]\ ,\nonumber\\
   &\epsilon^\nu_{\rm 2PK}=- 3 (\bar\beta_A + \bar\beta_B)+ \frac{1}{3} \bar\gamma_{AB}(10 +  \bar\gamma_{AB}) +\frac{1}{3} (\epsilon_A + \epsilon_B) + \frac{2}{3} (\delta_A + \delta_B) - 
      2 \zeta\ ,\nonumber
\end{align}
(\ref{AplusCorrections}) simply reads
\begin{equation}
A(u)=A^{\rm GR}_{2\rm PN}(u;\nu)+2\epsilon_{\rm 1PK} u^2+(\epsilon^0_{\rm 2PK}+\nu\,\epsilon^\nu_{\rm 2PK}) u^3\ .
\label{A1PKnonPade}
\end{equation}
Therefore, the Scalar-Tensor 2PK corrections to the $g^e_{00}$ component of the effective general relativistic 2PN metric are completely described by three parameters, $(\epsilon_{\rm 1PK}, \epsilon^0_{\rm 2PK}, \epsilon^\nu_{\rm 2PK})$, that are numerically of the same order of magnitude (since they are driven by $({\alpha_{A/B}^0})^2$).\\

When Scalar-Tensor effects are to be considered as perturbative,  our result (\ref{A1PKnonPade}) can be refined by replacing $A^{\rm GR}_{2\rm PN}(u;\nu)$ by the currently best available General Relativity EOB results, to which we turn now.

\subsection{ST-parametrised EOB dynamics}\label{sectionSTEOBNR}

We now propose to evaluate the effect of these ST post-Keplerian corrections to the general relativistic predictions for the ISCO frequency. To do so, we do not restrict ourselves to the 2PN GR expression for $A_{2\rm PN}^{\rm GR}(u ; \nu)$ but use instead the best available EOB-NR function $A^{\rm GR}(u ; \nu)$ (in the nonspinning case) :
 \begin{equation}
 A^{\rm GR}(u ; \nu)=\mathcal P^1_5[A^{Taylor}_{\rm 5PN}]\ ,
 \end{equation}
i.e. the $(1,5)$ Pad\'e approximant of the truncated 5PN expansion
\begin{equation}
A^{Taylor}_{\rm 5PN}=1-2u+2\nu u^3+\nu a_4u^4+(a_5^c+a_5^{\rm \ln}\ln u)u^5+\nu(a_6^c+a_6^{\rm \ln}\ln u)u^6\ ,
\label{AtaylorRG}
\end{equation}
where $a_6^c(\nu)$ has been obtained by calibration with Numerical Relativity results, the other coefficients being known analytically, see \cite{Damour:2014sva}, \cite{Nagar:2015xqa}, and \cite{Bini:2013zaa} for their explicit expressions.
Comparing (\ref{A1PKnonPade}) to (\ref{AtaylorRG}), Scalar-Tensor effects are clearly seen to induce a quadratic $\mathcal O (u^2)$ term that does not exist in General Relativity, and a ($\nu$-dependent) correction to the cubic $\mathcal O (u^3)$ coefficient.

One could in principle phenomenologically anticipate ST corrections coming from higher PK orders. However, first, it is known from General Relativity that from 3PN order on, the effective dynamics can not be that of a pure geodesic anymore (as mentioned in footnote \ref{footnote3PN}). Second, the two-body 3PK Lagrangian is not known in Scalar-Tensor theories. We hence leave these questions to future work and, for the time being, content ourselves with the study of the ST 2PK corrections only.

\vfill\eject

The study of the dynamics is now straightforward. By staticity and spherical symmetry of the metric (\ref{geodEffDroste}),
\begin{align}
\quad u_t=-A\frac{dt}{d\lambda}\equiv-E\ ,\quad u_\phi=r^2\frac{d\phi}{d\lambda}\equiv L\ ,\label{conservEandL}
\end{align}
are the conserved energy and angular momentum of the orbit, per unit mass $\mu$.
One also normalizes the 4-velocity $u^\alpha u_\alpha=-\epsilon$,
where $\epsilon=1$ for $\mu\neq 0$, $\epsilon=0$ for null geodesics ($\mu=0$).
The radial motion in the metric (\ref{geodEffDroste}) is hence determined by~:\begin{align}
&\left(\frac{dr}{d\lambda}\right)^2=\frac{1}{AB}F(u)\ ,\\
\text{where}\quad  F(u)\equiv E^2-A(u)&\left(\epsilon+j^2u^2\right)\quad,\quad j\equiv\frac{L}{G_{AB}M}\quad,\quad u\equiv\frac{G_{AB}M}{r}\ .
\end{align}
In the following we focus on circular orbits, assuming  that gravitational radiation has suppressed any eccentricity during the early inspiral.\footnote{Note that for circular orbits, motion can still be considered as geodesic at 3 and higher PN orders in General Relativity, see \cite{Damour:2000we}.\\} 

When $\epsilon=1$, the radial velocity vanishes when $F(u)=0$, while the circularity of the orbit also requires $F'(u)=0$ ; hence $j^2$ and $E$ are related to $u$ by :
\begin{equation}
j^2(u)=-\frac{A'}{(Au^2)'}\ ,\quad
E(u)=A\sqrt{2u\over (Au^2)'}\ .\label{Ej}
\end{equation}
The innermost stable circular orbit (``ISCO") requires the third (inflection point) condition $F''(u)=0$, i.e. $u_{\rm ISCO}$ is the root of the equation :
\begin{equation}
F'(u_{\rm ISCO})=F''(u_{\rm ISCO})=0\quad\Rightarrow\quad\frac{A''}{A'}=\frac{(Au^2)''}{(Au^2)'}\ .\label{ISCO}
\end{equation}
(As anticipated in the previous subsection  the circular orbits are determined by the function $A(u)$ only.)\\ 

Let us now turn to the real two-body dynamics. The quadratic relation between the real and effective Hamiltonians $H$ and $H_e$ (III.28) can be inverted to yield the EOB Hamiltonian, see (III.30)~:
\begin{equation}
H_{\rm EOB}=M\sqrt{1+2\nu\left(\frac{H_e}{\mu}-1\right)}\ ,\quad\text{where}\quad\frac{H_e}{\mu}=\sqrt{A\left(1+{\hat p_r^2\over B}+{\hat p_\phi^2\over \hat r^2}\right)}\ ,
\label{inverseRelationHam}
\end{equation}
which defines a resummed two-body dynamics.
Since $H_{\rm EOB}$ and $H_e$ are conservative, we have on-shell :
\begin{equation}
\left(\frac{\partial H_{\rm EOB}}{\partial H_e}\right)=\frac{1}{\sqrt{1+2\nu(E-1)}}\quad\text{since}\quad H_e=\mu E\quad\text{on-shell}\ .\label{timeRescaling}
\end{equation}
Hence the real (two-body problem) equations of motion have been drastically simplified, since  they now read
\begin{equation}
\frac{dr}{dt}=\frac{\partial H_{\rm EOB}}{\partial p_r}\ ,\quad\frac{d\phi}{dt}=\frac{\partial H_{\rm EOB}}{\partial p_\phi}\ ,\quad\frac{dp_\rho}{dt}=-\frac{\partial H_{\rm EOB}}{\partial r}\ ,\quad\frac{dp_\phi}{dt}=-\frac{\partial H_{\rm EOB}}{\partial\phi}=0\ ,
\end{equation}
and are identical to the effective equations of motion, to within the \textit{constant} multiplicative factor (\ref{timeRescaling}), i.e. a simple time rescaling $t\rightarrow t\sqrt{1+2\nu(E-1)}$.
Consequently, the effective orbital frequency (deduced from Hamilton's equations, or equivalently from (\ref{conservEandL})) being given by
\[
\omega(u)\equiv \frac{d\phi}{dt}=\frac{\partial H_e}{\partial p_\phi}={ju^2 A\over G_{AB}ME}\ ,
\]
the real frequency, deduced from $H_{\rm EOB}$, is
\begin{equation}
\Omega(u)=\frac{\partial H_{\rm EOB}}{\partial H_e}\frac{\partial H_e}{\partial p_\phi}=\frac{ju^2A}{G_{AB}ME\sqrt{1+2\nu(E-1)}}\ ,
\label{Omega}
\end{equation}
where $E(u)$ and $j(u)$ are given for circular orbits in (\ref{Ej}).\footnote{The orbital frequency has been derived in the ``effective", Droste, coordinate system, $(q,p)$. The real coordinates, $(Q,P)$, are linked to $(q,p)$ through the canonical transformation (\ref{transfoCano}-\ref{generatrice}) so that $\Phi\neq\phi$ in general. However, $\Phi=\phi$ for circular orbits ($p_r=P_R=0$) and hence (\ref{Omega}) is the real, observed, orbital frequency. Indeed, only the radial coordinates differ $r\neq R$, but are not observables. See section \ref{sectionTransfoCano}.\\}

\vfill\eject

Figure \ref{figureISCOeps} below shows the ISCO location in Droste coordinates and associated frequency when $\nu=1/4$ in the following cases :\\ \\
(i) considering only 1PK corrections, i.e. keeping only the $\mathcal O (u^2)$, $\epsilon_{\rm 1PK}$ term in (\ref{A1PKnonPade}). Note that $\epsilon_{1PK}$ can be negative. For instance, for identical bodies, $\epsilon_{1PK}=({\alpha_A^0})^2\left(\frac{1}{2}\beta_A^0+2\right)+\mathcal O({\alpha_A^0}^4)$ is driven by an overall $({\alpha_A^0})^2$ factor but is negative when $\beta_A^0<-4+\mathcal O(({\alpha_A^0})^2)$ ;\\ \\
(ii) adding 2PK, $\mathcal O (u^3)$ corrections. As discussed above, in Scalar-Tensor theories the 2PK coefficients are expected to be of the same order ; we hence incorporate them and, for simplicity, limit ourselves to the specific example~:
$$\epsilon_{\rm 2PK}^0+\nu\,\epsilon_{\rm 2PK}^\nu\equiv\epsilon_{\rm 1PK}$$
in the equal-mass case ($\nu=1/4$).

In both cases, the ISCO location and frequency are seen to increase dramatically as soon as $\epsilon_{\rm 1PK}$ approaches $\sim 10^{-1}$. What is happening here is similar to what was discussed at 3PN order in General Relativity in \cite{Damour:2000we} : when $\epsilon_{\rm 1PK}$ becomes too large and positive, the function $A(u)$ is no longer a good representation of the Scalar-Tensor deformations, since in particular, $A(u)$ has no zero anymore (in particular, it does not exhibit a horizon). This phenomenon is another reason to recall that this effective geodesic should be taken seriously only when Scalar-Tensor corrections are to be considered as perturbative (here, $\epsilon_{\rm 1PK}<<1$).\\

(iii) For that reason, we  follow the suggestion of \cite{Damour:2000we} and further resum $A(u)$ through an overall Pad\'e approximant, by continuity with the General Relativity ($\epsilon_{\rm 1PK}=\epsilon_{\rm 2PK}^0=\epsilon_{\rm 2PK}^\nu=0$) limit :
\begin{equation}
A^{\rm 2PK}(u)\equiv \mathcal P^1_5[A^{Taylor}_{\rm 5PN}+2\epsilon_{\rm 1PK} u^2+
(\epsilon_{\rm 2PK}^0+\nu\,\epsilon_{\rm 2PK}^\nu)u^3]\ ,\label{2PKpade}
\end{equation}
ensuring also that $A(u)$ has a simple zero. As one can see from Figure 1, the divergences are then efficiently cured.\footnote{See \cite{Barausse:2009xi} for a different resummation method.}\\

The ISCO frequency is roughly linear in $\epsilon_{\rm 1PK}$. The slope, or ``sensitivity" of the ISCO frequency to Scalar-Tensor corrections is
\begin{equation}
\left.\frac{d (G_{AB}M\Omega)_{ISCO}}{d\epsilon_{\rm 1PK}}\right|_{\nu=1/4}\simeq 0.13\ ,\quad\left.\frac{d (G_{AB}M\Omega)_{ISCO}}{d\epsilon_{\rm 1PK}}\right|_{\nu=0}\simeq 0.048\ .
\end{equation}

Finally, the relative correction reaches a few percents when $\epsilon_{\rm 1PK}\sim 10^{-2}$, see $x$ column in the Table below. It seems thus unlikely that measurements of this specific effect leads to improvements to the current (binary pulsar) constraints on Scalar-Tensor theories, (\ref{contrainteBinaryPulsars}).

\begin{figure}[h!]
\caption{Scalar-Tensor corrections to the ISCO location in Droste coordinates (left panel) and associated frequency (right panel) versus $\epsilon_{\rm 1PK}$ for $\nu=0.25$ and for $\epsilon_{\rm 2PK}^0+\nu\,\epsilon_{\rm 2PK}^\nu=\epsilon_{\rm 1PK}$. General Relativity is recovered when $\epsilon_{\rm 1PK}=0$. The first (dotted lines) and second (dashed lines) PK corrections quickly lead to divergences. The overall Pad\'e resummation (solid line) cures them efficiently. The table gathers a few numerical values in the 2PK Pad\'e resummed case ; $x\equiv G_{AB}M \Omega_{ISCO}/(G_{AB}M\Omega_{ISCO})_{GR}$.\label{tableLSO_2}}
\centering
\begin{subfigure}{.5\textwidth}
  \centering
  \includegraphics[width=0.95\linewidth]{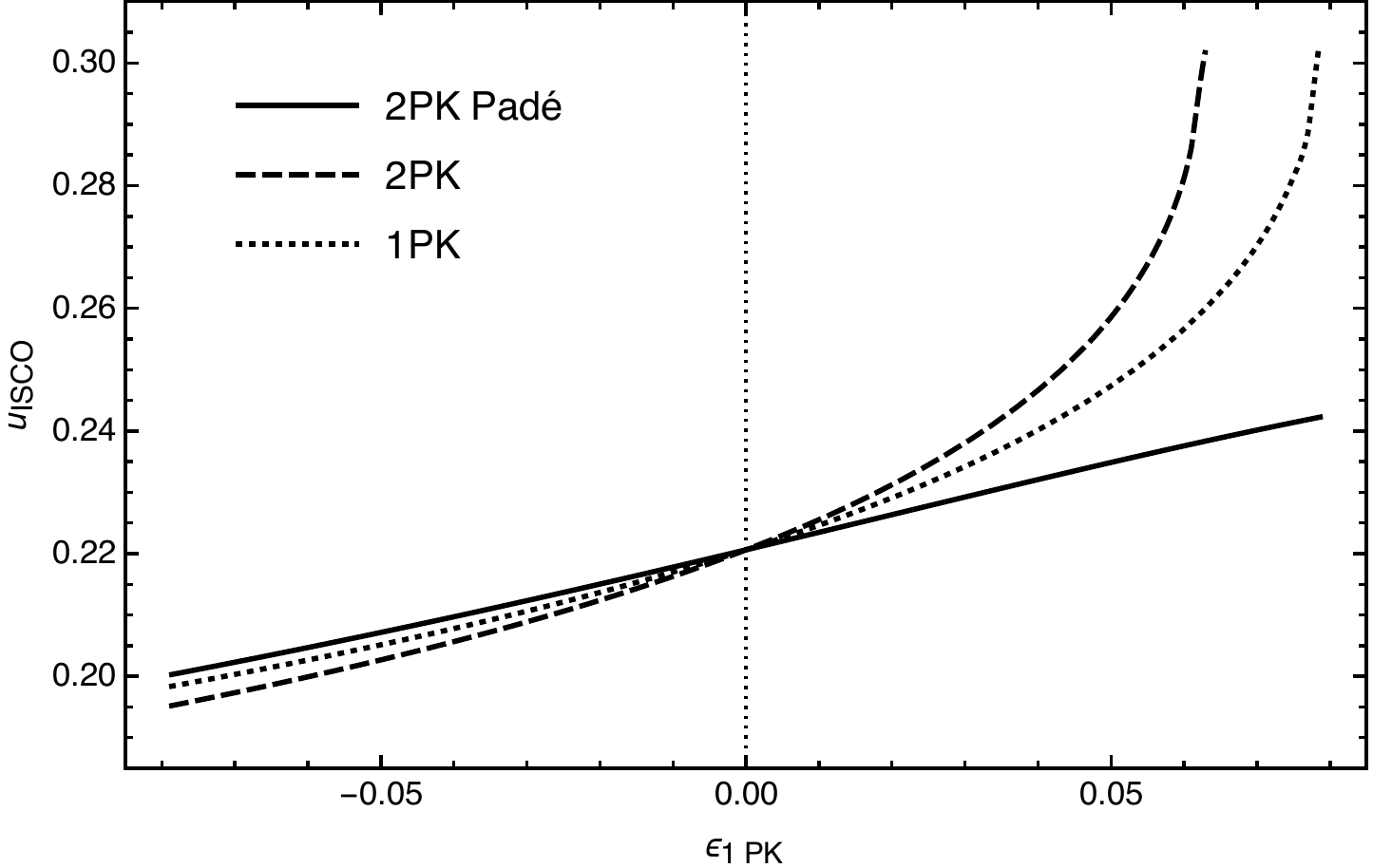}
  \label{fig:sub11}
\end{subfigure}%
\begin{subfigure}{.5\textwidth}
  \centering
  \includegraphics[width=0.95\linewidth]{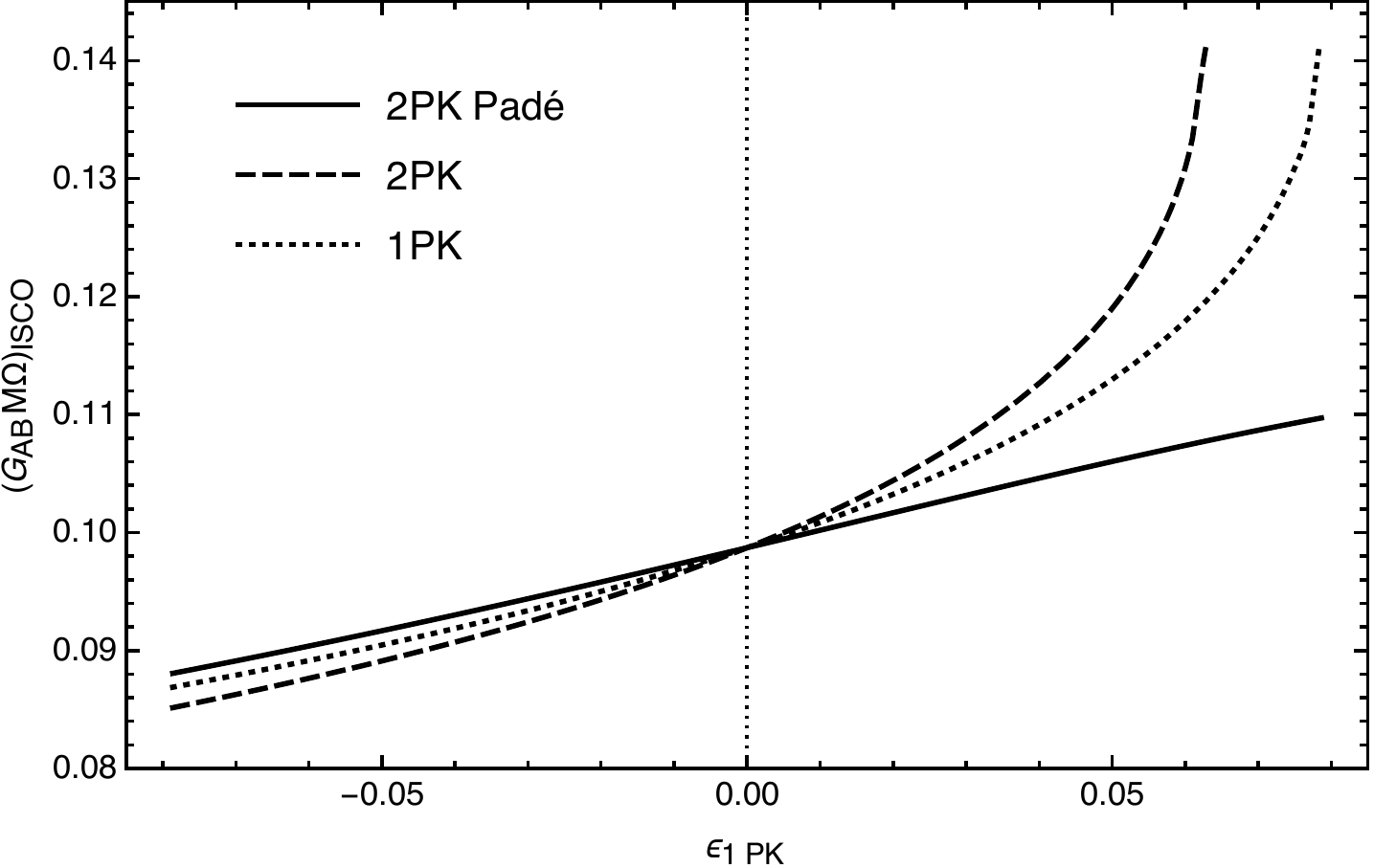}
  \label{fig:sub21}
\end{subfigure}
\label{figureISCOeps}
\end{figure}

\begin{table}[h!]
   \centering
   \renewcommand{\arraystretch}{1.2}
\begin{tabular}{| l | c c c | l | c c c |}
  \hline
  $\epsilon>0$ & $u_{ISCO}$ & $G_{AB}M\Omega_{ISCO}$ & $x$ & $\epsilon<0$ & $u_{ISCO}$ & $G_{AB}M\Omega_{ISCO}$ & $x$ \\
  \hline
  $10^{-3}$ & $0.2209$& 0.09886&1.002 &$-10^{-3}$ & 0.2203&0.09857 &0.9985\\
  $2.5\times 10^{-3}$ &0.2213 &0.09908 &1.004 &$-2.5\times 10^{-3}$ &0.2199 &0.09835 &0.9963\\
  $5\times10^{-3}$ & 0.2221&0.09945 &1.008 &$-5\times10^{-3}$ &0.2192 &0.09798 &0.9926\\
  $7.5\times10^{-3}$ &0.2228 &0.09982 &1.011 &$-7.5\times10^{-3}$ & 0.2185&0.09761 &0.9889\\
  $10^{-2}$ & 0.2235 &0.1002 &1.012 & $-10^{-2}$ &0.2178 &0.09725 &0.9852\\
  $2.5\times10^{-2}$ &0.2278 &0.1024 &1.038 & $-2.5\times10^{-2}$ &0.2137 &0.09510 &0.9634\\
  $5\times10^{-2}$ &0.2349 &0.1060 &1.074 &$-5\times10^{-2}$ &0.2072 &0.09168 &0.9287\\
  $7.5\times10^{-2}$ &0.2414 &0.1093 &1.107 &$-7.5\times10^{-2}$ &0.2011 &0.08851 &0.8966\\
  \hline
\end{tabular}
\end{table}



The study of circular geodesics in the metric (\ref{geodEffDroste}) has allowed us to describe the impact of the 2PK ST deviations to General Relativity (\ref{2PKpade}) on the ISCO frequency. In fact, as discussed in section \ref{EOBdroste}, any theory whose two-body Lagrangian verifies the constraints (IV.8-10) may also be mapped towards an effective geodesic. This suggests, by extension of the ST results, that (\ref{2PKpade}) takes the generic parametrized form :
\begin{equation}
A^{\rm PEOB}(u)\equiv \mathcal P^1_5[A^{Taylor}_{\rm 5PN}+2(\epsilon^0_{\rm 1PK}+\nu\,\epsilon^\nu_{\rm 1PK})u^2+(\epsilon^0_{\rm 2PK}+\nu\,\epsilon^\nu_{\rm 2PK}) u^3]\ ,
\end{equation}
where $\epsilon^0_{\rm 1PK}$, $\epsilon^\nu_{\rm 1PK}$, $\epsilon^0_{\rm 2PK}$, and $\epsilon^\nu_{\rm 2PK}$ are now to be regarded as theory-agnostic Parametrized EOB (PEOB) coefficients, and is suitable to encompass the (conservative) dynamics of a generic deviation to General Relativity at 2PK order. We note that no Keplerian parameter is needed since it can always be absorbed by a redefinition of the total mass (see, for example, $G_{AB}$ in the ST case). For Scalar-Tensor theories, $\epsilon_{\rm 1PK}^\nu=0$ and $\epsilon_{\rm 1PK}^0\sim\epsilon^0_{\rm 2PK}\sim\epsilon^\nu_{\rm 2PK}$.

\section{Concluding remarks}
It is a remarkable fact that the  EOB approach can be extended beyond the framework of General Relativity : the two-body (2PK) problem has indeed been mapped here towards the geodesic of an effective metric in Schwarschild-Droste coordinates. This paper is (to our knowledge) the first EOB description of a modified gravity, in the simplest example of massless Scalar-Tensor theories. \\

This mapping has led to a much simpler and compact (still, canonically equivalent) description of the two-body conservative dynamics in the 2PK regime, parlty hiding some of the irrelevant information of its Hamiltonian in an appropriate canonical transformation.
The effective problem also defines a resummation of the two-body dynamics that may capture some of its strong field features, in particular concerning the ISCO frequencies. In a second paper (in preparation), we shall build \textit{another} EOB Hamiltonian that maps the two-body problem to a $\nu$-deformed version of the Scalar-Tensor one-body problem.\\

The General Relativity EOB approach has been extended in \cite{Damour:2009wj} to the case of binary neutron stars. There, tidal effects were phenomenologically included by adding corrections to the $-g_{00}^e=A(u)$ part of the effective (Droste) metric, starting at 5PN order, i.e. $\mathcal O (u^6)$ (TEOB).
In contrast, our work should be regarded as a different extension, towards parametrized Scalar-Tensor theories (PEOB), and modifies the effective metric at 1PK order already, i.e.  $\mathcal O  (u^2)$ in $A(u)$. When applied to neutron stars, the Scalar-Tensor corrections must be compared to tidal effects. Our model shows that the PEOB $\mathcal O (u^2)$ corrections are generically numerically much smaller than the TEOB $\mathcal O (u^6)$ correction close to the merger, assuming the constraint $(\alpha_{A/B}^0)^2<4\times 10^{-6}$ discussed in section \ref{sectionParametrizedEOB}. However, systems that are subject to \textit{dynamical} scalarization \cite{Barausse:2012da} may develop nonperturbative scalar charges in the strong field regime, and hence escape this constraint. In that case, the ISCO frequency can be significantly modified as soon as $(\alpha_{A/B}^0)^2\gtrsim 10^{-2}$, see figure \ref{tableLSO_2}.\\

When it comes to the question of binary black holes, it is well known that static black holes in the Scalar-Tensor theory we are considering here cannot carry scalar hair and reduce to the Schwarzschild solution. However, this may no longer be true in the strong field, dynamical regime (i.e. near merger) which is \textit{precisely} explored by the EOB approach. 
Moreover, scalar hair can be induced by means of a potential $V(\varphi)$ or massless gauge fields. We leave the investigation of such effects to further work.\\

It should also be noted that while Solar System and binary pulsar observations have put stringent constraints on Scalar-Tensor theories, gravitational wave detectors are designed to detect highly redshifted sources, that is at cosmological epochs when Scalar-Tensor effects may have been more manifest (see e.g. \cite{Damour:1992kf} or \cite{Damour:1993id}). Therefore gravitational wave astronomy should be regarded as an opportunity to constrain also the cosmological history of Scalar-Tensor theories.\\

Finally, we restricted ourselves in this paper to the conservative part of the dynamics of the Scalar-Tensor two-body problem. The corresponding EOB radiation reaction force and gravitational waveforms still remain to be investigated and will be the topic of further work.

\section*{Acknowledgements}
We are very grateful to Thibault Damour for  introducing us to the subtleties of the EOB approach, for enlightning discussions and stimulating suggestions, all shortcomings being ours. FLJ also thanks Gilles Esposito-Far\`ese for his encouragement at an early stage of this project and for sharing his expertise in Scalar-Tensor theories.

\vfill\eject
\appendix

\section{Einstein vs Jordan frame - Conversion of the two-body parameters\label{appendixEvsJ}}
\begin{table}[h!]
\caption{Conversion of the two-body Lagrangian parameters\label{WillvsDEF}}
   \centering
   \renewcommand{\arraystretch}{1.7}
\begin{tabular}{|c l c|}
  \hline
   MW \cite{Mirshekari:2013vb} & DEF \cite{Damour:1992we}, \cite{Damour:1995kt}& This paper \\
  \hline
  \hline
  \textbf{Scalar-Tensor parameters}& &\\
  \hline
  $G$ & $\mathcal A_0^2(1+\alpha_0^2)$ & -\\
  $\zeta$ & $\frac{\alpha_0^2}{1+\alpha_0^2}$ & -\\
  $\lambda_1$ & $\frac{1}{2}\frac{\beta_0}{1+\alpha_0^2}$ & -\\
  $\lambda_2$ & $-\frac{1}{2(1+\alpha_0^2)^2}\left(\frac{\beta'_0\alpha_0}{2}+\beta_0\alpha_0^2-2\beta_0^2\right)$ & -\\
  \hline
  \textbf{Self-gravity (``sensitivity") parameters}& &\\
  \hline
  $s_{A}$& $\frac{1}{2}-\frac{\alpha^0_{A}}{2\alpha_0}$ &-\\
  $s'_A$ & $ \frac{\beta^0_A}{4\alpha^2_0}-\frac{\alpha_A^0\beta_0}{4\alpha^3_0}$ & -\\
  $s''_A$ & $-\frac{1}{2\alpha_0}\left(\frac{\beta'^0_A}{4\alpha_0^2}-\frac{3\beta_A^0\beta_0+\alpha^0_A\beta'_0}{4\alpha^3_0}+\frac{3\beta_0^2\alpha_A^0}{4\alpha_0^4}\right)$ & -\\
  \hline
  \textbf{Two-body Lagrangian parameters}& &\\
  \hline
  \textbf{Kepler}& &\\
  $m_{1}$ & $m^0_{A}/\mathcal A_0$ & $ m^0_{A}/\mathcal A_0 \equiv \tilde{m}^0_{A}$ \\
  $m_{2}$ & $m^0_{B}/\mathcal A_0$ & $ m^0_{B}/\mathcal A_0 \equiv \tilde{m}^0_{B}$ \\
  $G\alpha$ & $(1+\alpha^0_A\alpha^0_B)\mathcal A^2_0\equiv G_{AB}\mathcal A_0^2$ & $G_{AB}/\mathcal A_0^2\equiv\tilde{G}_{AB}$ \\
  \textbf{1PK}& &\\
  $\bar{\gamma}$ & $-2\frac{\alpha^0_A\alpha^0_B}{1+\alpha^0_A\alpha^0_B}\equiv\bar{\gamma}_{AB}$ & $\bar{\gamma}_{AB}$ \\
  $\bar{\beta}_1$ & $\frac{1}{2}\frac{(\beta_A\alpha_B^2)_0}{(1+\alpha_A^0\alpha_B^0)^2}\equiv\bar{\beta}^A_{BB}$ & $\bar{\beta}_A$\\
  $\bar{\beta}_2$ & $\frac{1}{2}\frac{(\beta_B\alpha_A^2)_0}{(1+\alpha_A^0\alpha_B^0)^2}\equiv\bar{\beta}^B_{AA}$ & $\bar{\beta}_B$\\
  \textbf{2PK}& &\\
  $\bar{\delta}_1$ & $\frac{(\alpha_A^0)^2}{(1+\alpha_A^0\alpha_B^0)^2}$  & $\delta_A$ \\
  $\bar{\delta}_2$ & $\frac{(\alpha_B^0)^2}{(1+\alpha_A^0\alpha_B^0)^2}$  & $\delta_B$\\
  $\bar{\chi}_1$ & $-\frac{1}{4}\frac{(\beta'_A\alpha_B^3)_0}{(1+\alpha_A^0\alpha_B^0)^3}\equiv-\frac{1}{4}\epsilon^A_{BBB}$ & $-\frac{1}{4}\epsilon_A$\\
  $\bar{\chi}_2$ & $-\frac{1}{4}\frac{(\beta'_B\alpha_A^3)_0}{(1+\alpha_A^0\alpha_B^0)^3}\equiv-\frac{1}{4}\epsilon^B_{AAA}$ & $-\frac{1}{4}\epsilon_B$\\
  $\bar{\beta}_1\bar{\beta}_2/\bar{\gamma}$ & $-\frac{1}{8}\frac{\beta^0_A\alpha^0_A\beta^0_B\alpha^0_B}{(1+\alpha_A^0\alpha_B^0)^3}\equiv-\frac{1}{8}\zeta_{ABAB}$ & $-\frac{1}{8}\zeta$\\
  \hline
\end{tabular}
\end{table}

In this appendix we convert the parameters appearing in the two-body (harmonic) Lagrangian of \cite{Mirshekari:2013vb} using the conventions introduced in \cite{Damour:1992we}. The Scalar-Tensor action reads, in the Einstein-frame (see section \ref{introduction}) :
\begin{equation}
S_{\rm EF}=\frac{1}{16\pi}\int d^4x\sqrt{-g}\bigg(R-2g^{\mu\nu}\partial_\mu\varphi\partial_\nu\varphi\bigg)+S_{m}\left[\Psi,\mathcal A^2(\varphi)g_{\mu\nu}\right]\ ,
\end{equation}
while in the conventions of \cite{Mirshekari:2013vb}, the action is written in the Jordan-frame as~:
\begin{equation}
S_{\rm JF}=\frac{1}{16\pi}\int d^4x\sqrt{-\tilde{g}}\left(\phi \tilde{R}-\frac{\omega (\phi)}{\phi}\tilde{g}^{\mu\nu}\partial_\mu\phi\partial_\nu\phi\right)+S_{m}\left[\Psi,\tilde{g}_{\mu\nu}\right]\ .
\end{equation} 
\vfill\eject
Hence, for a given function $\mathcal A(\varphi)$ charaterizing the ST theory in the Einstein-frame, the Jordan metric and function $\omega(\phi)$ characterizing it in the Jordan-frame are given by~: 
\begin{equation}
 \tilde{g}_{\mu\nu}=\mathcal A^2g_{\mu\nu}\ ,\quad \alpha=\frac{d\ln \mathcal A(\varphi)}{d\varphi}\ , \quad 3+2\omega(\phi)=\alpha(\varphi)^{-2}\,.\label{lienWillDEF}
\end{equation}
where $\varphi(\phi)$ is obtained by inverting $\mathcal A(\varphi)=1/\sqrt{\phi}$.
The parameters defined in table 1 of \cite{Mirshekari:2013vb} are translated using (\ref{JFmassEF}), (\ref{lienWillDEF}) and are gathered in table \ref{WillvsDEF}. In particular, we note that $\varphi(\phi_0)=\varphi_0$ are the background cosmological values of the scalar fields.

The notations of this paper are given in the third column. Some of them are a slight simplification of the Damour-Esposito Far\`ese parameters. Our table of correpondence agrees with \cite{Sennett:2016klh}, except for $\lambda_1$, $\lambda_2$, $s_A'$, $s_A''$. However this has no consequence on the two-body Lagrangian parameters, that we found to be in full agreement.\\

\section{The contact transformations defining the class of reduced Lagrangians\label{appendixTransfoContact}}
In section \ref{sectionElimTheAccel}, we performed a 2PK position redefinition (through a contact transformation) depending on the 14 parameters $f_i$ of the function $f$ introduced in (\ref{functionF}). Its full expression is :

\begin{align*}
&\delta\vec{Z}_A=\frac{G_{AB}m_B^0}{8}\left[2(7+4\bar\gamma_{AB})\vec{V}_B(\vec{N}\cdot\vec{V}_B)-\vec{N}\left((7+4\bar\gamma_{AB})V_B^2-(\vec{N}\cdot\vec{V}_B)^2\right)\right]\\
&-G_{AB}m_B^0\left[\vec V_A\bigg(2f_1(N\cdot V_A)-2f_4(N\cdot V_B)\bigg)+\vec{V}_B\bigg(f_2(N\cdot V_A)-f_5(N\cdot V_B)\bigg)\right.\\
&\left.+\vec N \bigg(f_1V_A^2+f_2V_A\cdot V_B+f_3V_B^2+3f_7(N\cdot V_A)^2+2f_8(N\cdot V_A)(N\cdot V_B)-f_9(N\cdot V_B)^2+f_{11}\frac{G_{AB}m_A^0}{R}+f_{12}\frac{G_{AB}m_B^0}{R}\bigg)\right]\ ,\end{align*}
\begin{align*}
&\delta\vec{Z}_B=\frac{G_{AB}m_A^0}{8}\left[-2(7+4\bar\gamma_{AB})\vec{V}_A(\vec{N}\cdot\vec{V}_A)+\vec{N}\left((7+4\bar\gamma_{AB})V_A^2-(\vec{N}\cdot\vec{V}_A)^2\right)\right]\\
&-G_{AB}m_A^0\left[\vec V_A\bigg(f_2(N\cdot V_A)-f_5(N\cdot V_B)\bigg)+\vec{V}_B\bigg(2f_3(N\cdot V_A)-2f_6(N\cdot V_B)\bigg)\right.\\
&\left.+\vec N \bigg(-f_4V_A^2-f_5V_A\cdot V_B-f_6V_B^2+f_8(N\cdot V_A)^2-2f_9(N\cdot V_A)(N\cdot V_B)-3f_{10}(N\cdot V_B)^2-f_{13}\frac{G_{AB}m_A^0}{R}-f_{14}\frac{G_{AB}m_B^0}{R}\bigg)\right]\ .
\end{align*}

\section{The two-body 2PK Hamiltonians for $f\neq 0$\label{appendixHamiltonien2body}}
When $f\neq 0$, we have on hands a whole class of ordinary Hamiltonians, corresponding implicitly to different coordinate systems. Only the 2PK coefficients differ from (\ref{coeffST2PK}) (because of the 2PK order contact transformations, see appendix \ref{appendixTransfoContact}) and read (see section \ref{sectionHamilt2bod})
$$
h^{\rm 2PK}_1=\frac{1}{16}\left(5 \nu^2-5 \nu+1\right),\quad h^{\rm 2PK}_2=h^{\rm 2PK}_3=h^{\rm 2PK}_4=0\ ,
$$

\begin{align*}
&h^{\rm 2PK}_5=\frac{G_{AB}}{8}\left[5  + 4 \bar{\gamma}_{AB} -( 22 + 16 \bar{\gamma}_{AB} ) \nu - 3  \nu^2\right]\\
 &+G_{AB}\left[- 
  f_6 \frac{m_A^0}{M} -  f_1 \frac{m_B^0}{M}+ 
   \nu\left( f_1 + f_6 + (f_1 + f_2 - f_4) \frac{m_B^0}{M} + (- f_3+f_5 + f_6 ) \frac{m_A^0}{M} \right)\right]\ ,
   \end{align*}
   \begin{align*}
   &h^{\rm 2PK}_6=G_{AB}\left[-\frac{\nu (\nu - 1)}{4}  + (f_1-3f_7) \frac{m_B^0}{M} + (f_6-3f_{10})\frac{m_A^0}{M}\right. \\ 
  &\hspace*{2cm}\left.-\nu \left( f_1 + f_6 - 3 f_7 - 
    3 f_{10} + \frac{m_B^0}{M} (f_1 + f_2 - f_4- 3 f_7 - 3 f_8 ) + \frac{m_A^0}{M} (-f_3 + f_5+ f_6 - 
       3 f_9  - 3 f_{10}) \right)\right]\ ,
       \end{align*}
       \begin{align*}
       &h^{\rm 2PK}_7=G_{AB}\left[-\frac{3}{8}\nu^2+3\left(  f_7 \frac{m_B^0}{M} +  f_{10} \frac{m_A^0}{M}\right) - 3
  \nu \left( f_7 + f_{10} + (f_7 + f_8) \frac{m_B^0}{M} + (f_9 + f_{10})\frac{m_A^0}{M}\right)\right]\ ,\end{align*}
  \begin{align*}&h^{\rm 2PK}_8=\frac{G_{AB}^2}{8}\left[ 22 - 4 \frac{m_A^0\bar\beta_B+m_B^0\bar \beta_A }{M} + 4 \frac{m_A^0\delta_A +m_B^0\delta_B }{M} + 28 \bar\gamma_{AB} + 
   9 \bar\gamma_{AB}^2 + \nu\left(58 - 4\frac{m_A^0\bar \beta_A+m_B^0\bar\beta_B}{M} + 36 \bar\gamma_{AB}\right)\right]\\
   &+G_{AB}^2\left[(f_1- f_{12}) \frac{m_B^0}{M} +(f_6- f_{13}) \frac{m_A^0}{M} + 
   \nu\left(- f_1-f_6- f_{11}+f_{12} +f_{13}  - f_{14}  - (f_1+f_2-f_4)  \frac{m_B^0}{M} + (f_3-f_5-f_6) \frac{m_A^0}{M}  \right)\right] ,
   \end{align*}
   \begin{align*}
   &h^{\rm 2PK}_9=G_{AB}^2\left[-\frac{1}{2} - \frac{1}{2} \frac{m_A^0 \delta_A + m_B^0 \delta_B}{M} - \frac{\bar\gamma_{AB}}{2} - \frac{\bar\gamma_{AB}^2}{8} +\nu \left( 
   -4 + (\bar\beta_A + \bar\beta_B) - 3 \bar\gamma_{AB} + \frac{m_A^0\bar\beta_B+m_B^0\bar\beta_A}{M}\right)\right]\\
   &+G_{AB}^2\left[(2 f_1+3 f_7+2f_{12} )\frac{m_B^0}{M} + (2 f_6 +3 f_{10}+2f_{13})\frac{m_A^0}{M}-\nu \bigg(2f_1+2 f_6+ 3 f_7 + 
  3 f_{10}-2f_{11}+2f_{12}+2f_{13}-2f_{14}  \right.\\
   &\hspace*{1cm}\left. +(2 f_1+ 2 f_2- 2 f_4+ 3 f_7+ 3 f_8) \frac{m_B^0}{M}  +( -2 f_3 + 2 f_5+2 f_6 + 3 f_9+ 
  3 f_{10})\frac{m_A^0}{M}\bigg)\right]\ ,
  \end{align*}
  \begin{align*}
  &h^{\rm 2PK}_{10}=G_{AB}^3
  \left[-\frac{1}{2}  - \frac{m_B^0\bar\beta_A+m_A^0\bar\beta_B}{M} - \frac{1}{6}\frac{m_A^0\epsilon_B+m_B^0\epsilon_A}{M}- \frac{1}{3}\frac{m_A^0\delta_A+m_B^0\delta_B}{M} - \frac{\bar\gamma_{AB}}{3}  - \frac{\bar\gamma_{AB}^2}{12}+f_{12}  \frac{m_B^0}{M} + f_{13} \frac{m_A^0}{M} \right.\\
  &\hspace*{2cm}+ 
   \nu \bigg(-\frac{15}{4} - \zeta + \frac{\bar\gamma_{AB}^2}{6} - \frac{4}{3} \bar\gamma_{AB} +  \frac{\delta_A + \delta_B}{3} + \frac{\epsilon_A+\epsilon_B}{6} -  (\bar\beta_A+\bar\beta_B)+f_{11}-f_{12}-f_{13}+f_{14}\bigg) \bigg]\ .
\end{align*}
They reduce to II.22 for $f=0$. As a consistency check, one retreives the General Relativistic ADM coordinates Hamiltonian (given e.g. in \cite{Buonanno:1998gg}), that is,  in the limit (\ref{limiteRG}),  setting :
\begin{equation}
f_3=f_4=-\frac{1}{4}\ ,\quad f_{12}=f_{13}=\frac{1}{4}\ ,\quad f_{11}=f_{14}=\frac{7}{4}\ ,
\label{limiteADM}
\end{equation}
the other $f_i$ coefficients being zero.

\section{Canonically-transformed two-body Hamiltonian\label{appendixPostTranfoCano}}
By means of a generic canonical transformation (\ref{transfoCano}-\ref{generatrice}), the two-body (2PK) Hamiltonian (see section \ref{sectionHamilt2bod}) is rewritten in the intermediate coordinate system $(Q,P)\rightarrow(Q,p)$ (recalling the notation ${\cal P}^2\equiv\hat p_r^2+{\hat p_\phi^2\over\hat R^2}$)~:
\begin{equation}
\hat H=\frac{M}{\mu}+\left(\frac{{\cal P}^2}{2}-\frac{h^{\rm K}}{\hat R}\right)+\hat H^{\rm 1PK}+\hat H^{\rm 2PK}+\cdots\ ,
\end{equation}
where
\begin{align*}
&\hat H^{\rm 1PK}=h^{\rm 1PK}_1 {\cal P}^4+{\cal P}^2 \hat p_r^2 \bigg(h^{\rm 1PK}_2-\alpha _1\bigg)+\hat p_r^4 \bigg(2 \alpha _1+\beta _1+h^{\rm 1PK}_3\bigg) + \frac{h^{\rm 1PK}_4 {\cal P}^2 + h^{\rm 1PK}_5 \hat p_r^2}{\hat R}+\frac{h^{\rm 1PK}_6}{\hat R^2}\ ,\\
&\hat H^{\rm 2PK}={\cal P}^2 \hat p_r^4 \bigg(-2 \alpha _1^2+4 \alpha _2-\beta _2+4 \beta _1 h^{\rm 1PK}_1+\alpha _1 \left(-\beta _1+8 h^{\rm 1PK}_1+2 h^{\rm 1PK}_2-4 h^{\rm 1PK}_3\right)+2 \beta _1 h^{\rm 1PK}_2+h^{\rm 2PK}_3\bigg)\\
&+\frac{1}{2} {\cal P}^4 \hat p_r^2 \bigg(\alpha _1^2-4 \alpha _1 (2 h^{\rm 1PK}_1+h^{\rm 1PK}_2)+2 \left(h^{\rm 2PK}_2-3 \alpha _2\right)\bigg)\ ,\\
&+\hat p_r^6 \left(2 \alpha _1^2+\frac{\beta _1^2}{2}+2 \beta _2+\gamma _2+2 \alpha _1 \left(\beta _1+2 h^{\rm 1PK}_2+4 h^{\rm 1PK}_3\right)+2 \beta _1 (h^{\rm 1PK}_2+2 h^{\rm 1PK}_3)+h^{\rm 2PK}_4\right)+h^{\rm 2PK}_1 {\cal P}^6\\
&+\frac{1}{\hat R}\bigg[{\cal P}^2 \hat p_r^2 \bigg(-2 \delta _2-2 \alpha _1 (h^{\rm 1PK}_4+h^{\rm 1PK}_5)+h^{\rm 2PK}_6\bigg)+\hat p_r^4 \bigg(2 \delta _2+4 \alpha _1 (h^{\rm 1PK}_4+h^{\rm 1PK}_5)+2 \beta _1 (h^{\rm 1PK}_4+h^{\rm 1PK}_5)+h^{\rm 2PK}_7\bigg)+h^{\rm 2PK}_5 {\cal P}^4\bigg]\\
& +\frac{1}{\hat R^2}\bigg[h^{\rm 2PK}_8 {\cal P}^2+\hat p_r^2 \bigg(h^{\rm 2PK}_9-\eta _2\bigg) \bigg]+\frac{h^{\rm 2PK}_{10}}{\hat R^3}\ .
\end{align*}
and where the 17 coefficients $h_i^{N{\rm PK}}$, which depend on the 14 parameters $f_i$, are given in Appendix C. 

\section{Canonically-transformed effective Hamiltonians\label{appendixCanonicalEffective Hamiltonians}}
Performing the canonical transformation (\ref{transfoCano}-\ref{generatrice}), the effective (2PK expanded) Hamiltonians are rewritten in the intermediate coordinate system $(q,p)\rightarrow(Q,p)$. The Keplerian order is unaffected by the canonical transformation, as discussed below (\ref{transfoCano}).\\

The Hamiltonians presented in section \ref{sectionEffectGeodDroste} read (recalling the notation ${\cal P}^2\equiv\hat p_r^2+{\hat p_\phi^2\over\hat R^2}$)
\begin{align*}
&\hat H_e^{1PN}=-\left(\alpha _1+\frac{1}{8}\right) {\cal P}^4-\hat p_r^2 {\cal P}^2 \bigg(\alpha _1+3 \beta _1\bigg)+\hat p_r^4 \bigg(2 \alpha _1+3 \beta _1\bigg)\\
&+\frac{1}{4\hat R}\bigg[{\cal P}^2 \bigg(a_1 \left(1-2 \alpha _1\right)-4 \gamma _1\bigg)-2 \hat p_r^2 \bigg(a_1 \left(2 \alpha _1+3 \beta _1\right)+b_1-2 \gamma _1\bigg)\bigg]+\frac{4 a_2-a_1 \left(a_1+4 \gamma _1\right)}{8 \hat R^2}\ ,
\end{align*}
\begin{align*}
&\hat H_e^{2PN}=\frac{1}{16} \bigg(24 \alpha _1^2+8 \alpha _1-16 \alpha _2+1\bigg) {\cal P}^6+\frac{1}{2} \hat p_r^2 {\cal P}^4 \bigg(\alpha _1 \left(18 \beta _1+1\right)+9 \alpha _1^2-6 \alpha _2+3 \beta _1-6 \beta _2\bigg)\\
&+\frac{1}{2} \hat p_r^4 {\cal P}^2 \bigg(2 \alpha _1 \left(9 \beta _1-1\right)+8 \alpha _2+27 \beta _1^2-3 \beta _1+2 \beta _2-10 \gamma _2\bigg)-\frac{1}{2} \hat p_r^6 \bigg(36 \alpha _1 \beta _1+12 \alpha _1^2+27 \beta _1^2-4 \beta _2-10 \gamma _2\bigg)\\
&+\frac{1}{16\hat R}\bigg[8 \hat p_r^4 \bigg(a_1 \left(2 \alpha _1 \left(6 \beta _1+1\right)+4 \alpha _1^2+9 \beta _1^2+3 \beta _1-2 \beta _2-5 \gamma _2\right)-12 \alpha _1 \gamma _1+b_1 \left(2 \alpha _1+3 \beta _1\right)-18 \beta _1 \gamma _1+4 \delta _2+6 \epsilon _2\bigg)\\
&+4 \hat p_r^2 {\cal P}^2 \bigg(a_1 \left(4 \alpha _1 \left(3 \beta _1-1\right)+8 \alpha _1^2-8 \alpha _2-9 \beta _1-6 \beta _2\right)+2 \left(\gamma _1 \left(6 \alpha _1+18 \beta _1-1\right)-2 \left(\delta _2+3 \epsilon _2\right)\right)+\left(2 \alpha _1+1\right) b_1\bigg)+\\
&{\cal P}^4 \bigg(a_1 \left(8 \alpha _1^2-12 \alpha _1-8 \alpha _2-1\right)+8 \left(\left(6 \alpha _1+1\right) \gamma _1-2 \delta _2\right)\bigg)\bigg]\\
&+\frac{1}{16\hat R^2}\bigg[4 \hat p_r^2 \bigg(a_1^2 \left(2 \alpha _1+3 \beta _1\right)-a_1 \left(b_1-2 \left(\gamma _1 \left(4 \alpha _1+6 \beta _1+1\right)-2 \delta _2-3 \epsilon _2\right)\right)+2 \left(-4 a_2 \alpha _1-6 a_2 \beta _1+b_1 \gamma _1+b_1^2-b_2\right.\\
&\left.-3 \gamma _1^2+2 \eta _2\right)\bigg)
+{\cal P}^2 \bigg(4 a_1 \left(\left(4 \alpha _1-3\right) \gamma _1-2 \delta _2\right)+4 \left(a_2 \left(1-4 \alpha _1\right)+6 \gamma _1^2-4 \eta _2\right)+\left(4 \alpha _1-1\right) a_1^2\bigg)\bigg]\\
&+\frac{1}{16 \hat R^3}\bigg[-4 a_1 \left(a_2-2 \gamma _1^2+2 \eta _2\right)+4 a_1^2 \gamma _1+8 \left(a_3-2 a_2 \gamma _1\right)+a_1^3\bigg]\ .
\end{align*}

\vskip1.8cm

\bibliographystyle{unsrt}
\bibliography{biblio}

\end{document}